\documentclass[letterpaper,aps,prd,twocolumn,tightenlines,preprintnumbers,nofootinbib,showkeys,superscriptaddress,longbibliography]{revtex4-1}

\usepackage{lipsum} %%%%%%%%%%%%%%%%%%%%%%%%%%%%%%%%%%%%%%%%%%%%%%%%%%%%%%%%%%%%%%%%%%

\usepackage{XCharter}
\usepackage[T1]{fontenc}
\usepackage{bm}
\usepackage{bbold}
\usepackage{pifont}
\usepackage{rotating}
\usepackage{fullpage}
\usepackage{amsfonts,shorthand}
\usepackage{amsmath}
\usepackage{slashed}
\usepackage{siunitx}
\usepackage{amssymb}
\usepackage{graphicx}
\usepackage{epic}
\usepackage{eepic}
\usepackage{epsfig}
\usepackage{latexsym}
\usepackage[dvipsnames]{xcolor}
\usepackage[export]{adjustbox}
\usepackage{float}
\usepackage{multirow, makecell}
\usepackage{hyperref}
\usepackage{enumitem}
\hypersetup{colorlinks=true,citecolor=red,linkcolor=NavyBlue,urlcolor=NavyBlue}
\usepackage[caption=false]{subfig}
 
\usepackage{natbib}
\usepackage{relsize}
\usepackage[left=1.65cm,right=1.65cm,top=1.83cm,bottom=1.84cm]{geometry}
\begin{document}
\relscale{1.05}
%%%%%%%%%%%%%%%%%%%%%%%%%%%%%%%%%%%%%%%%%%
\title{Machine learning-enhanced search for a vectorlike singlet $\mathbf B$ quark decaying to a singlet scalar or pseudoscalar}
%%%%%%%%%%%%%%%%%%%%%%%%%%%%%%%%%%%%%%%%%%
\author{Jai Bardhan}
\email{jai.bardhan@research.iiit.ac.in} 
\affiliation{Center for Computational Natural Sciences and Bioinformatics, International Institute of Information Technology, Hyderabad 500 032, India}

\author{Tanumoy Mandal}
\email{tanumoy@iisertvm.ac.in}
\affiliation{Indian Institute of Science Education and Research Thiruvananthapuram, Vithura, Kerala, 695 551, India}

\author{Subhadip Mitra}
\email{subhadip.mitra@iiit.ac.in}
\affiliation{Center for Computational Natural Sciences and Bioinformatics, International Institute of Information Technology, Hyderabad 500 032, India}

\author{Cyrin Neeraj}
\email{cyrin.neeraj@research.iiit.ac.in}
\affiliation{Center for Computational Natural Sciences and Bioinformatics, International Institute of Information Technology, Hyderabad 500 032, India}
%%%%%%%%%%%%%%%%%%%%%%%%%%%%%%%%%%%%%%%%%%
\date{\today}
%%%%%%%%%%%%%%%%%%%%%%%%%%%%%%%%%%%%%%%%%%
\begin{abstract}
\noindent 
The presence of a new decay mode relaxes the current mass exclusion limits on vectorlike quarks considerably. We consider the case of a weak-singlet vectorlike $B$ quark that can decay to a singlet scalar or pseudoscalar $\Phi$. In an earlier paper~[A. Bhardwaj \emph{et al.}, Roadmap to explore vectorlike quarks decaying to a new scalar or pseudoscalar, \href{https://doi.org/10.1103/PhysRevD.106.095014}{{\it Phys. Rev.} D, {\bf 106} (2022) 095014}; \href{http://arxiv.org/abs/2203.13753}{arXiv:2203.13753}], we mapped the possibilities to explore such setups at the LHC. We showed that it is possible for a $B$ quark to decay into $\Phi$ and the $\Phi$ to dominantly decay to a pair of gluons or $b$ quark(s) without fine-tuning the parameters. In this paper, we present a collider search strategy to look for the pair production of singlet $B$ quarks. If both $B$'s decay into $b\Phi$ pairs, the final state is fully hadronic: $B{B}\to(b\Phi)({b}\Phi)\to (bgg)({b}gg)/(bb{b})({b}b{b})$, which is very challenging to probe. Therefore, we consider a simpler mixed decay specific to the singlet $B$ case, $BB\to(b\Phi)(tW)$ with a lepton in the final state, to achieve the best sensitivity at the high-luminosity LHC. We use a deep neural network with a weighted cross-entropy loss to separate the signal from the huge SM background. Our analysis shows that large areas of the $M_{B}-M_{\Phi}$ parameter space are discoverable through this signature. We show how the discovery and exclusion regions scale with the branching ratio in the new decay mode. We also estimate the reach in the inclusive monolepton channel with the same network model.
\end{abstract}
%%%%%%%%%%%%%%%%%%%%%%%%%%%%%%%%%%%%%%%%%%
\maketitle

%%%%%%%%%%%%%%%%%%%%%%%%%%%%%%%%%%%%%%%%%%
\section{Introduction}
\noindent 
Extensions of the Standard Model (SM) involving vectorlike quarks (VLQs) are frequently invoked to explain various experimental observations or discrepancies. Generally, these extensions are, from a top-down point of view, well-motivated and can successfully address some of the shortcomings of the SM~\cite{Kaplan:1983fs,Kaplan:1991dc,Agashe:2004rs,Ferretti:2013kya,Gherghetta:2000qt,Contino:2003ve,Arkani-Hamed:2002iiv,Arkani-Hamed:2002ikv}. Depending on their quantum numbers, VLQs can decay to the SM particles by mixing with the SM quarks. The mixing is generally with the third-generation quarks as those are comparatively less constrained by the flavour-changing neutral-current data. Looking for VLQs exclusively decaying to third-generation quarks and heavy bosons ($W/Z/H$) is a major search program at the LHC. The prospects of these channels have been extensively studied before (see, e.g., Refs.~\cite{Gopalakrishna:2011ef,Dermisek:2019vkc,Choudhury:2021nib,Banerjee:2016wls}). However, in many new-physics theories, an extended quark sector with VLQs naturally appears along with an extended scalar sector containing heavy scalar/pseudoscalar particles~\cite{Gopalakrishna:2013hua,Erdmenger:2020lvq,Erdmenger:2020flu,Barcelo:2014kha}. In some cases, a VLQ can also decay to a new scalar or pseudoscalar~\cite{Gopalakrishna:2015wwa,Serra:2015xfa,Kraml:2016eti,Dobrescu:2016pda,Aguilar-Saavedra:2017giu,Chala:2017xgc,Moretti:2017qby,Bizot:2018tds,Colucci:2018vxz,Han:2018hcu,Anandakrishnan:2015yfa,Xie:2019gya,Benbrik:2019zdp,Wang:2020ips,Dermisek:2021zjd,Dermisek:2020gbr,Corcella:2021mdl,Dasgupta:2021fzw}, or dark photons~\cite{Kim:2019oyh,Verma:2022nyd}, often significantly~\cite{Cacciapaglia:2019zmj,Bhardwaj:2022nko}. The null results in the current LHC searches motivate us to seriously investigate such exotic decay modes of the VLQs.

In Ref.~\cite{Bhardwaj:2022nko}, we presented a 
roadmap to search for vectorlike top/bottom partners in the presence of a lighter weak-singlet scalar or pseudoscalar ($\Phi$) at the LHC. Even though the $\Phi$ has no direct couplings with the SM quarks, they are generated  through the mixing of VLQs ($Q$) with the corresponding third-generation quarks ($q$) after electroweak-symmetry breaking (EWSB). As a result, a new VLQ-decay mode opens up: $Q\to q\Phi$. A singlet $\Phi$ can decay to digluon ($gg$), diphoton ($\gm\gm$) and diboson ($VV^\prime$) states through the VLQ loops. The $q\leftrightarrow Q$ mixing also allows the $\Phi$ to have a $q\bar{q}$ decay at the tree level. As charted out in Ref.~\cite{Bhardwaj:2022nko}, the presence of an additional decay mode for the VLQs and the subsequent decay of the $\Phi$ opens a host of new and interesting possibilities in a large region of parameter space without any fine-tuning. 

In a follow-up paper~\cite{Bhardwaj:2022wfz}, we studied the prospects of the pair production of $T$ quark decaying exclusively to the new boson, $pp\to T\bar T\to t \Phi \bar t \Phi$, with each $\Phi$ decaying into two gluon jets at the high-luminosity LHC (HL-LHC). In this paper, we look at the case of the $B$ quark. There are a few major differences between these two cases. For example, in the case of $B\to b\Phi$, the $\Phi$ dominantly decays to two jets, either two gluons or two $b$ quarks, in the entire parameter region of interest (see Ref.~\cite{Bhardwaj:2022nko} for the other possible modes).  As a result, when the new decay mode of $B$ dominates, the resultant final states are purely hadronic, i.e., $B\bar{B}\to \left(b g g\right) \left(\bar{b} g g\right) / \left(b b \bar{b}\right) \left(\bar{b} b \bar{b}\right)$. This is unlike the case of $T$, where the final states have at least one top quark---one can use various analysis strategies (like, tagging a boosted top quark, or looking for a leptonically decaying top, etc.) depending on the top quark decay modes to identify such a topology. The absence of a top quark in the fully-hadronic final states makes the searches for the $pp\to B\bar B\to b\Phi\bar b\Phi$ mode challenging.
%%%%%%%%%%%%%%%%%%%%%%%%%%%%%%%%%%%%%%%%%%
\begin{figure}[t!]
    \centering
    \includegraphics[width=0.375\textwidth]{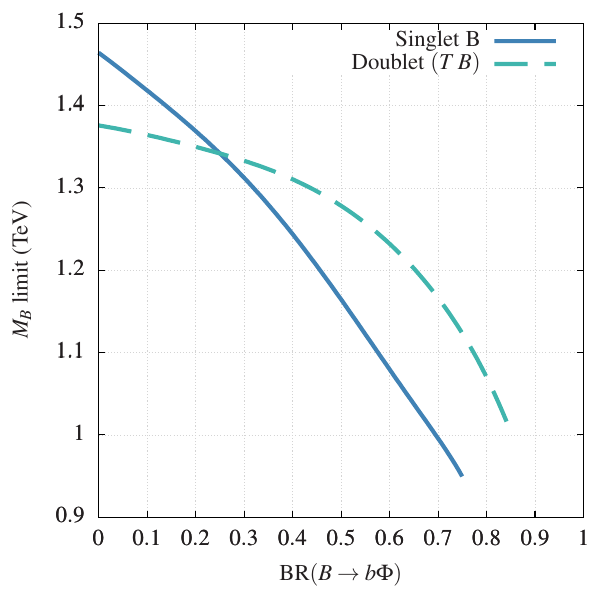}
    \caption{LHC exclusion limits on $B$ in the singlet $B$ and doublet model as a function of the branching ratio in the new decay mode.}
    \label{fig:b2_LHClimits}
\end{figure}
%%%%%%%%%%%%%%%%%%%%%%%%%%%%%%%%%%%%%%%%%%

Even though the LHC is yet to search for a $B$ decaying to a singlet scalar, it is possible to draw bounds on the possible scenarios from the LHC data. In Ref.~\cite{Bhardwaj:2022nko}, we recast the current VLQ searches~\cite{CMS:2020ttz, CMS:2019eqb, ATLAS:2022hnn} to put mass exclusion bounds on $B$ in the presence of a new decay mode by adapting the branching ratio (BR) condition as
\begin{equation}
\label{eq:brEq}
\beta_{bH} + \beta_{bZ} + \beta_{tW} = 1\to \left(1 - \beta_{b\Phi}\right)
\end{equation}
where $\beta_{X}=$ BR$(B \to X)$. When the mass of the $B$ is more than a TeV, $\beta_{t W} \approx 2\beta_{b Z} \approx 2\beta_{b H}$ if the $B$ is a weak-singlet (the singlet $B$ model in Ref.~\cite{Bhardwaj:2022nko}), and $ \beta_{b Z} \approx \beta_{b H},\ \beta_{t W}\approx 0$ if it is component of a $(T\ B)$ doublet (the doublet  model). We show the recast limits for the two possible $B$ representations in Fig.~\ref{fig:b2_LHClimits}. [In addition to the rescaled bound on VLQs,  we also draw the bounds on the square of the $\Phi\to gg$ coupling times BR$(\Phi\to\gm\gm)$ from the ATLAS heavy-resonance search data in the diphoton mode~\cite{ATLAS:2021uiz}
in the same paper. Constraints from the measurement of $Z$ boson coupling to the left-handed $b$ quark~\cite{Gopalakrishna:2013hua, Agashe:2006at} and flavour-changing neutral-current couplings~\cite{Nir:1990yq} restrict the $b \leftrightarrow B$ mixing from being arbitrarily large.

From the rescaled $B$ limits in Fig.~\ref{fig:b2_LHClimits}, we see another difference from the $T$ case. The limits are not as restrictive as they are for the $T$---the mass limits relax significantly, especially for the singlet $B$ with $\bt_{b\Phi}\gtrsim 0.2$. This can also be read as a relaxing condition on the BR in the extra mode. A smaller BR in the new mode implies larger BRs in the SM modes. Hence, in addition to the challenging fully hadronic final states, one can also make use of the semileptonic final states to search for the $B$ quark, like $pp\to B\bar B\to b\Phi\, \bar tW^+ + \bar b\Phi\, tW^-$ where the $t$ or the $W$ decays leptonically, especially since identifying a lepton is much easier at the LHC (which can be used as a trigger).

From the above mixed-decay modes of $B\bar{B}$, we see that two kinds of signal topologies are possible: (a) one with exactly one lepton and the other with (b) two leptons. The monolepton signature is exclusive to the singlet $B$ model as the $tW$ mode is negligible in the doublet model~\cite{Bhardwaj:2022nko}. The dilepton signature is possible when a $B$ decays to a leptonically-decaying $Z$ boson. Hence it is common to both singlet and doublet models. However, to suppress the huge Drell-Yan dilepton background, a $Z$-veto cut is necessary. This cut not only suppresses the background but also kills the signal.

%%%%%%%%%%%%%%%%%%%%%%%%%%%%%%%%%%%%%%%%%%
\begin{figure}[t!]
    \centering
    \includegraphics[width=0.375\textwidth]{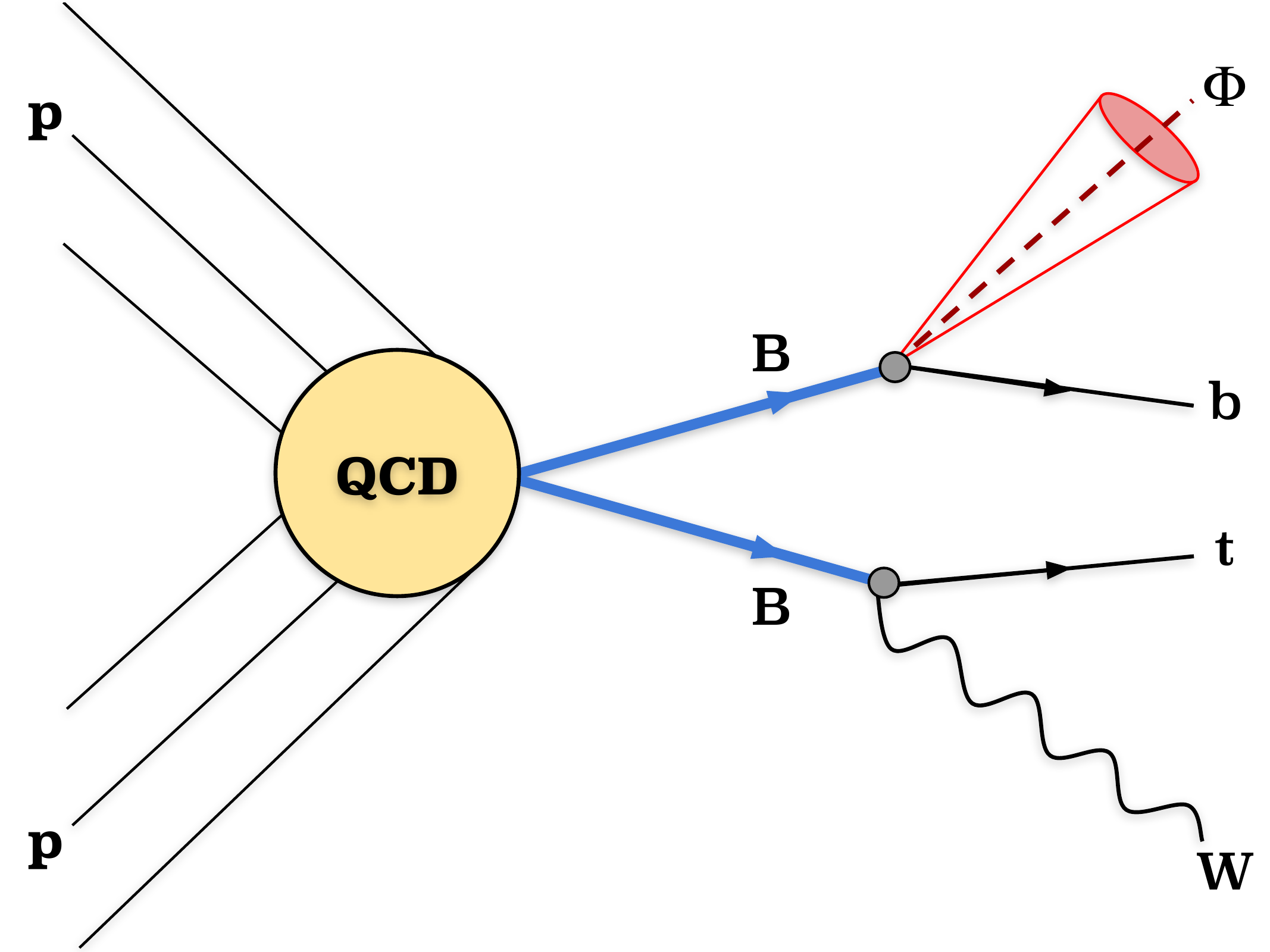}
    \caption{Signal topology}
    \label{fig:FeynDiag}
\end{figure}
%%%%%%%%%%%%%%%%%%%%%%%%%%%%%%%%%%%%%%%%%%
In this paper, we investigate the prospects of the mixed-decay mode with monolepton final states (see Fig.~\ref{fig:FeynDiag}) at the HL-LHC since it is an exclusive possibility for a close-to-TeV $B$. For the singlet $B$ model, this relatively simpler channel might have good prospects compared to the purely hadronic ones. We use a deep neural-network (DNN) model to isolate the signal. Using the same deep-learning model, we also present the collider reach for a pair produced singlet $B$ which decays to at least one $W$ boson and at least one top quark (i.e., $BB \to tW + X$) producing one lepton in the final state.

The rest of the paper is organised as follows. We describe the singlet $B$ model in Sec.~\ref{sec:model}; discuss the collider phenomenology of the model and define the event selection criteria in Sec.~\ref{sec:colliderpheno}; describe the neural network (NN) we use in Sec.~\ref{sec:dnn} and discuss the event selection criteria for the inclusive channel in Sec.~\ref{sec:IncMode}. Finally, we present the results in Sec.~\ref{sec:results} and conclude in Sec.~\ref{sec:conclu}.
%%%%%%%%%%%%%%%%%%%%%%%%%%%%%%%%%%%%%%%%%%
\begin{figure*}[htp!]
\captionsetup[subfigure]{labelformat=empty}
\subfloat[\quad\quad(a)]{\includegraphics[width=0.75\columnwidth]{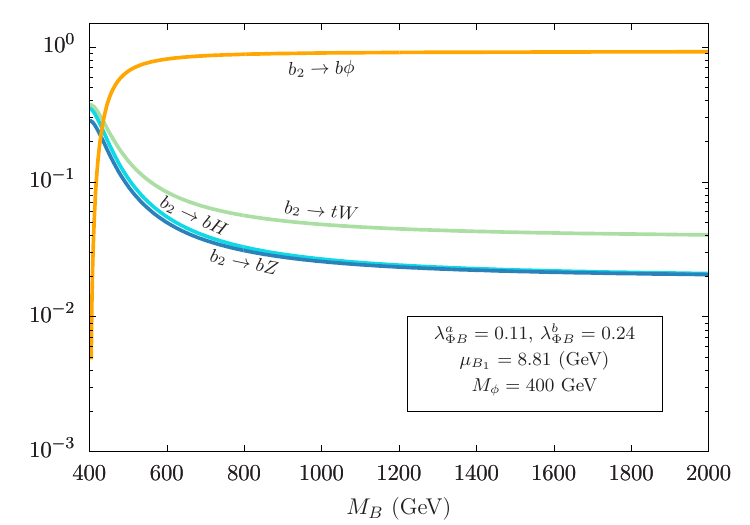}\label{fig:br_B_bMark1}}\hspace{0.5cm}
\subfloat[\quad\quad(b)]{\includegraphics[width=0.75\columnwidth]{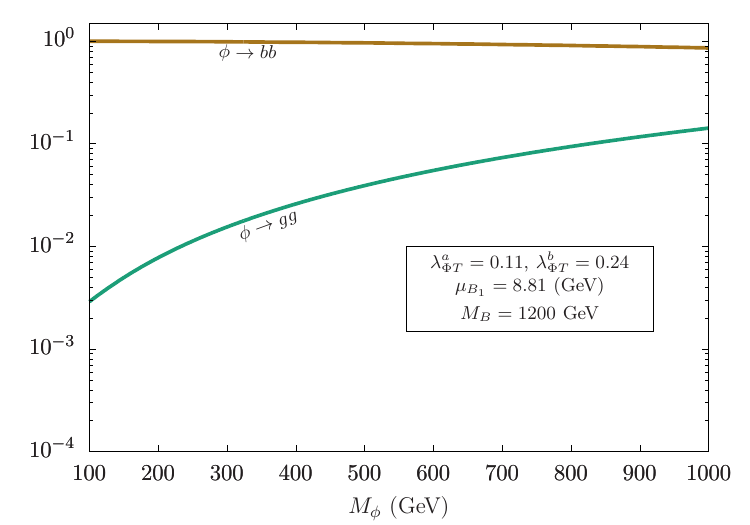}\label{fig:br_Phi_bMark1}}\\
\subfloat[\quad\quad(c)]{\includegraphics[width=0.75\columnwidth]{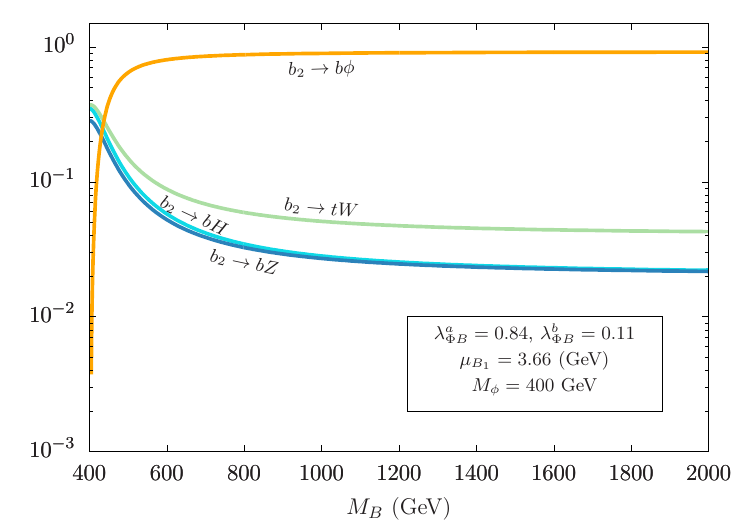}\label{fig:br_B_bMark2}}\hspace{0.5cm}
\subfloat[\quad\quad(d)]{\includegraphics[width=0.75\columnwidth]{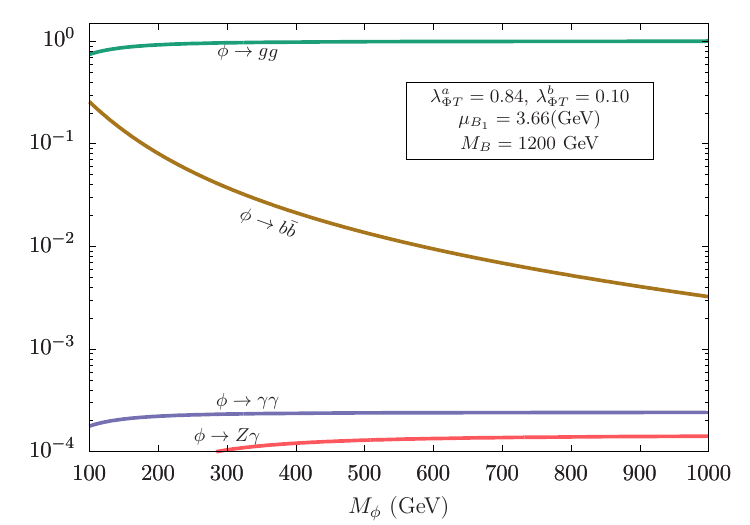}\label{fig:br_Phi_bMark2}}\\
\subfloat[\quad\quad(e)]{\includegraphics[width=0.75\columnwidth]{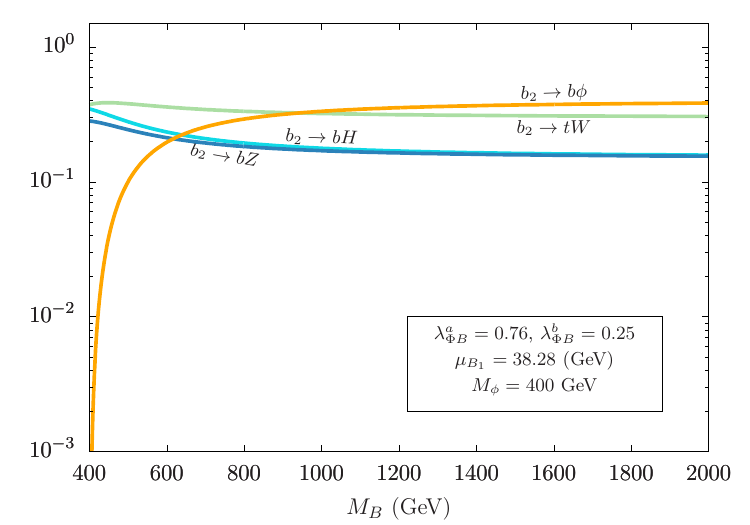}\label{fig:br_B_bMark3}}\hspace{0.5cm}
\subfloat[\quad\quad(f)]{\includegraphics[width=0.75\columnwidth]{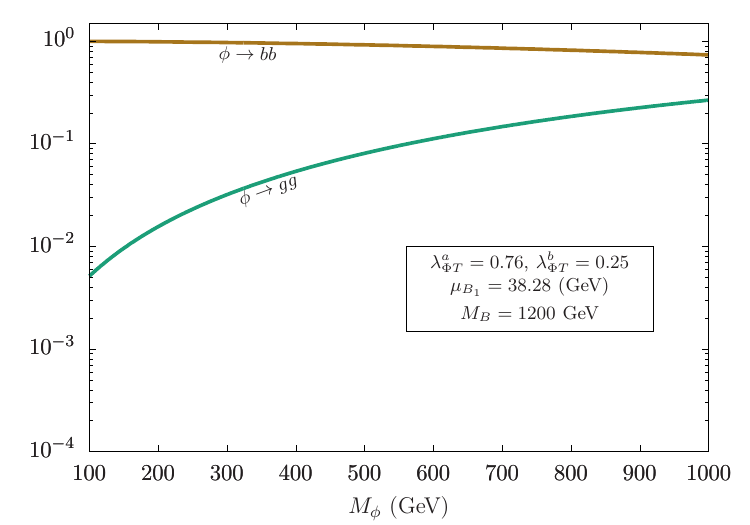}\label{fig:br_Phi_bMark3}}\\
\subfloat[\quad\quad(g)]{\includegraphics[width=0.75\columnwidth]{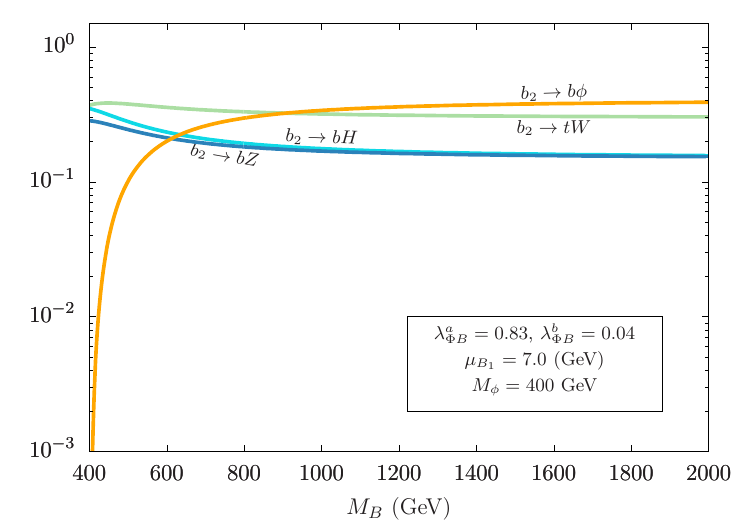}\label{fig:br_B_bMark4}}\hspace{0.5cm}
\subfloat[\quad\quad(h)]{\includegraphics[width=0.75\columnwidth]{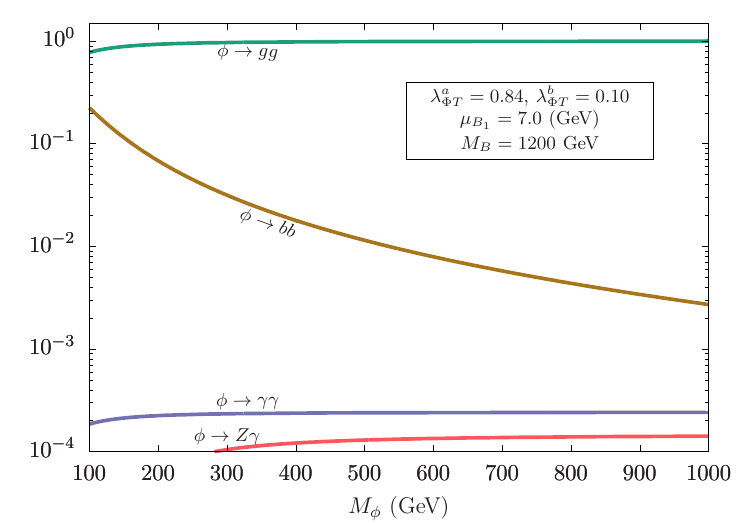}\label{fig:br_Phi_bMark4}}\\
\caption{Decays of $B$ and $\phi$ in the singlet $B$ model for benchmark choice of parameters. The $B\to b\phi$ decay dominates over all the other decay modes in (a)--(d) whereas BR$(B\to tW)\approx$ BR$(B\to b\phi)$ in (e)--(h). On the right panel, we show that either $\phi\to gg$ or $\phi\to b\bar{b}$ can dominate in both scenarios.}
\label{fig:br}
\end{figure*}
%%%%%%%%%%%%%%%%%%%%%%%%%%%%%%%%%%%%%%%%%%

%%%%%%%%%%%%%%%%%%%%%%%%%%%%%%%%%%%%%%%%%%
%%%%%%%%%%%%%%%%%%%%%%%%%%%%%%%%%%%%%%%%%%
\begin{figure*}
\captionsetup[subfigure]{labelformat=empty}
\centering
\subfloat[\quad\quad(a)]{\includegraphics[width=0.375\textwidth]{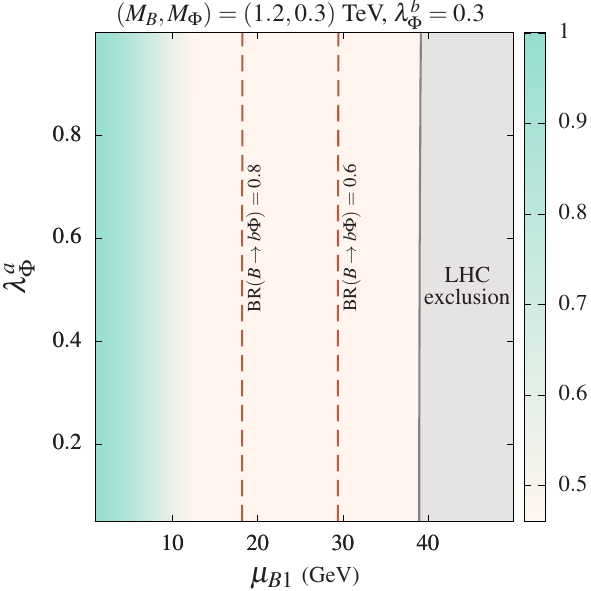}\label{fig:lamBconst}}\hspace{1cm}
\subfloat[\quad\quad(b)]{\includegraphics[width=0.375\textwidth]{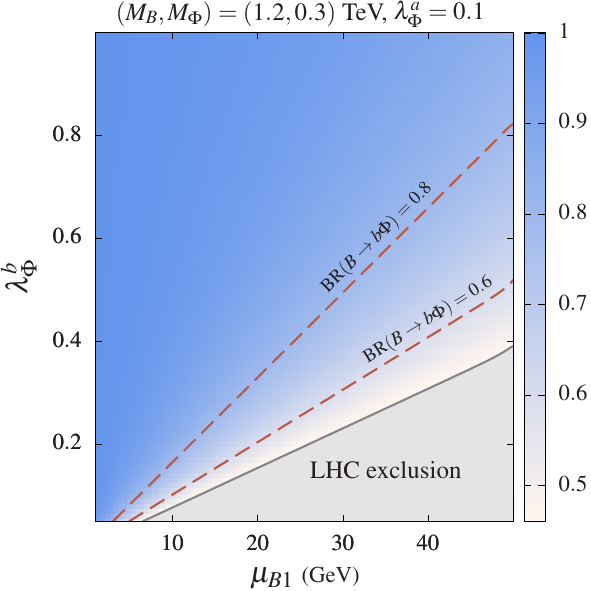}\label{fig:lamAconst}}
\caption{Branching-ratio maps (contours) in terms of the model parameters for  $(M_{B}, M_{\Phi}) = (1.2, 0.3)$ TeV. Here, $\m_{B1} = \om_B v/\sqrt 2$ is the off-diagonal term in the mass matrix.}
\label{fig:BRcontours_ModelParam}
\end{figure*}
%%%%%%%%%%%%%%%%%%%%%%%%%%%%%%%%%%%%%%%%%%
\section{Singlet $B$ Model}\label{sec:model}
\noindent
We consider the  singlet $B$ model from Ref.~\cite{Bhardwaj:2022nko} which is a simple extension of the SM containing a TeV-scale weak-singlet $B\equiv\left(\mathbf{3},\mathbf{1}, -1/3\right)$ and a sub-TeV weak-singlet scalar or pseudoscalar $\Phi\equiv\left(\mathbf{1},\mathbf{1}, 0\right)$. In a similar notation, the mass terms relevant to the bottom sector can be written as
\begin{align}
    \mathcal{L} \supset -\Big\{&\lambda_{b}\left(\bar{Q}_LH_B\right)b_R
    +\omega_{B}\left(\bar{Q}_LH_B\right)B_R\ \nonumber\\
    & + M_{B}\bar{B}_LB_R+ h.c.\Big\},\label{eq:massmat}
\end{align}
where $Q_L$ is the third-generation quark doublet and $H_B^T=1/\sqrt{2} \begin{pmatrix} 0 & v \end{pmatrix}$, with $v$ being the vacuum expectation value of the Higgs field. The off-diagonal Yukawa coupling is denoted by $\lambda_b$, $\omega_B$ denotes the mixing between the SM and the vectorlike $B$, and $M_{B}$ is the mass term of $B$. After EWSB, we get the mass matrix $\mathcal{M}$ as,
\begin{equation}
    \mathcal{L}_{mass} = \begin{pmatrix} \bar{b}_L & \bar{B}_L \end{pmatrix} 
    \begin{pmatrix}
    \lambda_b \frac{v}{\sqrt2} & \omega_{B}\,\frac{v}{\sqrt2} \\ 0 & M_B
    \end{pmatrix}
    \begin{pmatrix} b_R \\ B_R \end{pmatrix} + h.c.
\end{equation}
The matrix is diagonalised by a bi-orthogonal rotation with two mixing angles $\theta_L$ and $\theta_R$,
\begin{equation}
    \begin{pmatrix} b_{L/R} \\ B_{L/R}\end{pmatrix} = 
    \begin{pmatrix} c_{L/R} & s_{L/R} \\ -s_{L/R} & c_{L/R} \end{pmatrix}
    \begin{pmatrix} b_{1_{L/R}} \\ b_{2_{L/R}} \end{pmatrix},
\end{equation}
where $s_{P}=\sin\theta_{P}$ and $c_{P}=\cos\theta_{P}$ for the two chirality projections, and $b_1$ and $b_2$ are the mass eigenstates. We identify $b_1$ as the physical bottom quark and $b_2$ is essentially the $B$ quark for small $\om_B$. Hence, we use the notations $B$ and $b_2$ interchangeably. The Lagrangian for $\Phi$ interactions is given by
\begin{equation}
    \mathcal{L} = -\lambda^{a}_{\Phi}\Phi \bar{B}_L \Gamma B_R - \lambda_{\Phi}^{b}\Phi \bar{B}_L\Gamma b_R + h.c.,\label{eq:lambdas}
\end{equation}
where $\Gamma=\left\{1, i\gamma_{5}\right\}$ for $\Phi=\left\{ \mbox{scalar~}\phi, \mbox{pseudoscalar~}\eta \right\}$. Expanding in terms of $b_1, b_2$ the above Lagrangian gives,
\begin{align}
    \mathcal{L} = &-\lambda_{\Phi}^{a}\Phi\left(c_L \bar{b}_{2L} - s_L \bar{b}_L\right)\Gamma\left(c_R b_{2R} - s_R b_R\right) \nonumber \\
                        &-\lambda_{\Phi}^{b}\Phi\left(c_L \bar{b}_{2L} - s_L \bar{b}_L\right)\Gamma\left(c_R b_{R} + s_R b_{2R}\right) + h.c.
\end{align}

Figure 4 of Ref.~\cite{Bhardwaj:2022nko} shows that in the singlet $B$ model, the dominant decay mode of $\Phi$ is not determined by the choice of the $\lm$ couplings shown in the above equations. It rather depends on the value of the off-diagonal mass term $\mu_{B1}=\om_B v/\sqrt2$; the $\Phi\to gg$ mode dominates for small $\m_{B1}$ values. The $\Phi\to b\bar b$ mode starts dominating with increasing $\m_{B1}$. In Fig.~\ref{fig:br}, we show some benchmark choices of parameters and the corresponding decays of $B$ and the scalar $\phi$. The $B\to b\phi$ decay dominates over all the other decay modes in Figs.~\ref{fig:br_B_bMark1}--\ref{fig:br_Phi_bMark2} whereas BR$(B\to tW)\approx$ BR$(B\to b\phi)$ in Figs.~\ref{fig:br_B_bMark3}--\ref{fig:br_Phi_bMark4}. The same also holds true for a pseudoscalar $\eta$. To cover both possibilities, our analysis relies only on the two-pronged nature of $\Phi$ (i.e., no $b$-tagging) for a comprehensive study of the prospects of the singlet $B$ model in the monolepton channel. 

Relating the model parameters to the branching ratio in the new mode is straightforward, as explained in Ref.~\cite{Bhardwaj:2022nko}. For an illustration, we show the BR$(B\to b\Phi)$ contours for a benchmark point of $(M_{B},M_{\Phi}) = (1.2, 0.3)$ TeV  in Fig.~\ref{fig:BRcontours_ModelParam}.
%%%%%%%%%%%%%%%%%%%%%%%%%%%%%%%%%%%%%%%%%%
\section{Collider phenomenology}\label{sec:colliderpheno}
\begin{figure}[t!]
    \centering
    \includegraphics[width=0.375\textwidth]{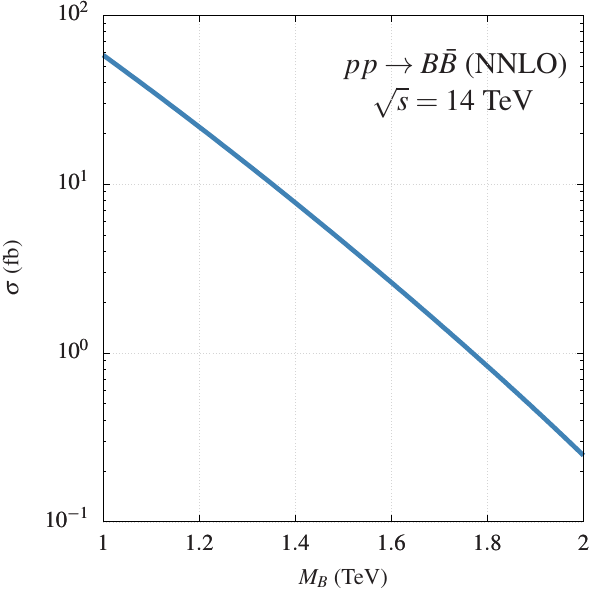}
\caption{The signal cross section at the $14$ TeV LHC calculated from the leading order cross section as $\sg$(LO)$\times K_{\rm NNLO}$. We estimate the NNLO QCD $K$-factor from Ref.~\cite{CMS:2019eqb} as $1.3$.}
    \label{fig:ppXS}
\end{figure}

\noindent 
We use \textsc{FeynRulesv2.3}~\cite{Alloul:2013bka} to build the singlet $B+\Phi$ model and obtain the {\sc Universal FeynRules Output}~\cite{Degrande:2011ua} model files. We use \textsc{MadGraph5v3.3}~\cite{Alwall:2014hca} to simulate the hard-scatterings at the leading order, \textsc{Pythia8}~\cite{Sjostrand:2014zea} for showering and hadronisation, and \textsc{Delphes3}~\cite{deFavereau:2013fsa} for simulating a generic LHC detector environment. The events are generated at $\sqrt{s} = 14$ TeV. To account for the boosts of the final state objects, we have modified the default \textsc{Delphes} card. For the electron, we update the \texttt{DeltaRMax} parameter (the maximum radius of an electron cone centred around the identified track) from $0.5$ to $0.2$, as per Ref.~\cite{ATLAS:2019qmc}. The $b$-tagging efficiency and the mistag rate for the lighter quarks were updated to reflect the medium working point of the DeepCSV tagger from Ref.~\cite{CMS:2017wtu}. We consider leptons with $p_T > 10$~GeV and $|\eta| < 2.5$. Our analysis relies on two types of jets, both clustered using the anti-$k_T$  algorithm~\cite{Cacciari:2008gp}, one with $R=0.4$ (AK-4) and the other with $R=1.2$ (fatjet). The AK-4 objects are required to pass a minimum-$p_T$ cut, $p_T > 20$ GeV and have $|\eta|<5$. We use the next-to-next-to-leading order (NNLO) signal cross sections (Fig.~\ref{fig:ppXS}) and for the background processes, we use the highest-order cross sections available in the literature (Table~\ref{tab:bg-crossx}).
%%%%%%%%%%%%%%%%%%%%%%%%%%%%%%%%%%%%%%%%%%
%%%%%%%%%%%%%%%%%%%%%%%%%%%%%%%%%%%%%%%%%%
%%%%%%%%%%%%%%%%%%%%%%%%%%%%%%%%%%%%%%%%%%
\begin{table}[t!]
\centering
\caption{Higher-order cross sections of the SM backgrounds considered in our analysis.\label{tab:bg-crossx}}
{\renewcommand\baselinestretch{1.25}\selectfont
\begin{tabular*}{\columnwidth}{l @{\extracolsep{\fill}} lrc}
\hline
Background                   &                   & $\sigma$           & QCD        \\
Processes                    &                   & (pb)               & order      \\
\hline \hline
\multirow{2}{*}{$V$ + jets~\cite{Balossini:2009sa, PhysRevLett.103.082001}} & $W$ + jets  & $1.95 \times 10^5$ & NLO        \\
                                                    & $Z$ + jets  & $6.33 \times 10^4$ & NNLO        \\
\hline
$tt$~\cite{Muselli:2015kba} & $tt$ + jets & $988.57$           & N$^3$LO    \\
\hline
\multirow{3}{*}{Single $t$~\cite{Kidonakis:2015nna}} & $tW$        & $83.10$            & N$^{2}$LO  \\
                             & $tb$              & $248.00$           & N$^{2}$LO \\
                             & $t$ + jets        & $12.35$            & N$^{2}$LO  \\
\hline
\multirow{2}{*}{$VV$ + jets~\cite{Campbell:2011bn}} & $WW$ + jets       & $124.31$           & NLO        \\
                             & $WZ$ + jets       & $51.82$            & NLO        \\
\hline                             
\multirow{3}{*}{$ttV$~\cite{Kulesza:2018tqz,LHCHiggsCrossSectionWorkingGroup:2016ypw}}       & $ttZ$             & $1.05$             & NLO + NNLL \\
                             & $ttW$             & $0.65$             & NLO + NNLL \\
                             % & $ttH$             & $0.6113$           & NLO        \\
                             & $ttH$             & $0.61$           & NLO       \\
\hline
\end{tabular*}}
\end{table}
%%%%%%%%%%%%%%%%%%%%%%%%%%%%%%%%%%%%%%%%%%

%%%%%%%%%%%%%%%%%%%%%%%%%%%%%%%%%%%%%%%%%%
\subsection{Signal topology}
\noindent 
The signal process we look for is $pp \to B\bar{B} \to \left(b\Phi\right)\:\left(\bar{t} W^{+}\right)$, where one of the $B$'s decays to a $b$ quark and a $\Phi$, the other to a $t$ quark and a $W$ boson. The $\Phi$ then decays to a pair of $b$ quarks or gluons [the parity of $\Phi$ does not affect our results, so we refer to the new boson as $\Phi$ even though we have used a scalar ($\phi$) to generate the signal events] and the $tW$ pair decays semileptonically, i.e., exactly one lepton comes out the decay of either the top quark or the $W$ boson. Thus, the essential topology of our signal can be summarised in terms of high-$p_T$ objects as
\begin{align}
\mbox{exactly } 1 \mbox{ lepton } + \mbox{ at least } 1~ b \mbox{ jet } + \mbox{ jets}.
\end{align}

To generate events for each signal benchmark, we pick model parameters such that BR$(B \to b\Phi) \approx 0.6$ while ensuring that narrow width approximation remains valid for both $B$ and $\Phi$. Thus, for the $M_{B}$--$M_{\Phi}$ parameter space that we scan over, the results can be easily scaled for any BR in the new mode. We choose our analysis to be independent of the $\Phi$ branching in $b\bar{b}$ or $gg$ mode by not explicitly ($b$-)tagging the decay products of $\Phi$. (However, we evaluate the analysis strategy on the events generated with $\Phi \to b\bar{b}$ and remark about the $gg$ mode analysis when it is relevant.)

%%%%%%%%%%%%%%%%%%%%%%%%%%%%%%%%%%%%%%%%%%
\subsection{Topology motivated selection criteria}
\noindent
As the mass difference between $B$ and $\Phi$ increases, the $\Phi$ can be tagged as a boosted fatjet. Looking at the signal topology, we design a set of selection criteria to select events to pass on to a DNN. We demand that each selected event must have the following characteristics:
\begin{itemize}\label{set:CutsExcl}
    \item[$\mathfrak{C}_1$:] \emph{Exactly 1 lepton} ($\ell \in \left\{e,\mu\right\}$). \\ The lepton is required to have a $p_T > 100$ GeV, $|\eta| < 2.5$ and must obey the updated isolation criteria mentioned earlier. 
    
    \item[$\mathfrak{C}_2$:] $H_T > 900$~GeV, where $H_T$ is the scalar sum of the transverse momenta of all hadronic objects in an event.
    
    \item[$\mathfrak{C}_3$:] \emph{At least 3 AK-4 jets with} $p_T > 60$~GeV. \\ The leading jet must have $p_T > 120$~GeV. 
    
    \item[$\mathfrak{C}_4$:] \emph{At least 1 b-tagged jet with} $p_T > 60$~GeV. \\ At least one of the AK-4 jets must be identified as a $b$ jet.
    
    \item[$\mathfrak{C}_5$:] \emph{At least 1 fatjet ($J$) with $R=1.2$ and} $p_T > 500$ GeV. \\ The fatjet is clustered using the anti-$k_T$ algorithm and the parameters have been optimised to tag a $\Phi$ fatjet. We also demand the invariant mass of the fatjet, $M_{J}$, to be more than $250$ GeV.
    
    \item[$\mathfrak{C}_6$:] \emph{$\Delta R_{b J} > 1.2$.}\\We demand that at least one of the identified $b$ jets is well separated from the leading fatjet passing $\mathfrak{C}_5$, i.e., the $b$ jet lies outside the fatjet cone. 
\end{itemize}
The invariant mass cut on the $\Phi$ fatjet limits contamination from the top or $W$ fatjets and the $b$ isolation criterion in $\mathfrak{C}_6$ reduces the background contribution from the semileptonic $t\bar t$ process significantly but does not affect the signal greatly (since the hardest $b$ comes from a $B$ quark in the signal). We set the mass range of $\Phi$ as $M_{\Phi} \in \left[ 300,700 \right]$ in accordance with the benchmark masses of $\Phi$ considered in Ref.~\cite{Bhardwaj:2022nko}. We see that for a given mass of $B$, the final efficiency after these cuts scales positively with $M_{\Phi}$---the fatjet criteria is more efficient at detecting $\Phi$ of higher masses.

%%%%%%%%%%%%%%%%%%%%%%%%%%%%%%%%%%%%%%%%%%
\subsection{Background processes}
\noindent 
We identify the SM background processes that can mimic the signal and project their contributions at the HL-LHC. 
\begin{itemize}
    \item[\ding{111}] \emph{$V+$ jets}, where  $V \in \left\{W^{\pm}, Z\right\}$ :
    A priori, $W^\pm (+ 2j)$ is the most dominant monolepton background of all the SM processes. While generating the events we ensure that the $W$-boson decays leptonically and up to two jets are matched with parton showers. This background is severely cut at the analysis stage due to the requirement of a $b$ jet as well as a high invariant mass fatjet. Furthermore, the requirement of high $p_T$ jets also cuts this background. We also consider leptonically decaying $Z (+ 2j)$, but this background falls considerably after the lepton, $b$-jet, and fatjet cuts.  
    
    \item[\ding{111}] Semileptonic $t\bar{t}+$  jets:   
    We simulate the process by matching up to two extra jets. Its topology closely matches our signal as a fatjet can come from the hadronically decaying top. However, the additional requirement of a reasonably well-separated $b$ jet from the fatjet significantly cuts down this process contribution. This is due to the fact that the separated $b$ jet has to come from the leptonically decaying $t$ quark.

    \item[\ding{111}] Fully leptonic $t\bar{t}$ + jets:
    We also consider the fully leptonic case after matching up to two extra jets for completeness' sake. This background is mitigated by the requirements on the fatjet.

    \item[\ding{111}] Diboson backgrounds ($VV +$ jets):
    Of the diboson backgrounds, we consider the ones that match the selection criteria: $W_{\ell}W_h$ and $W_{\ell}Z_h$. We simulate these processes by matching one extra jet. The $b$ jet and heavy fatjet requirements reduce these backgrounds.

    \item[\ding{111}] $t\bar{t}V + $ jets, where $V \in \left\{W^{\pm},Z,H\right\}$:
    It is similar to the semileptonic $t\bar{t}$ process with an additional SM boson. These processes mimic the signal topology well---more than the $t\bar{t}$ background. When $V = W^\pm$, the lepton can come from either of the top quarks or the $W$ boson. The leptonic mode of $Z$ has a dileptonic signature and is not considered. The reconstructed fatjet is generally the hadronic top quark, but $V$ may also be reconstructed. The requirement of $\Delta R_{bJ} > 1.2$ cuts out the $ttW^\pm$ background, but due to the significant branching of the $b\bar{b}$ mode of the $Z/H$ decay has a lesser effect on the processes containing them.   
    
    \item[\ding{111}] $tW^{\pm} + $ jets: 
    Here, either the $t$ quark or $W$ decays leptonically and the hadronically decaying object can form a fatjet. However, the $\Delta R_{b J} > 1.2$ requirement cuts the number of top fatjets.
    
    \item[\ding{111}] $t+b$/jet:
    We consider the leptonically decaying single-top process. Its contribution falls significantly after the kinematic cuts are enforced (e.g., Scalar $H_T$, $p_T$ cuts on the jets and the lepton, etc.).
\end{itemize}

\label{sec:gen-info}In order to save computation time, we generate these background processes with some generation-level cuts which are reinforced at the analysis level selection criteria. We summarise the effects of the cuts $\mathfrak{C}_1$--$\mathfrak{C}_6$ on the signal and background events in Table~\ref{tab:cut-flow}.

%%%%%%%%%%%%%%%%%%%%%%%%%%%%%%%%%%%%%%%%%%
\begin{table*}
\begin{centering}
\caption{Cut flow for the three benchmark choices of the signal and the relevant background processes. We use the MLM jet-parton shower matching technique~\cite{Mangano:2006rw} to generate the background samples as indicated by the additional jets in brackets. The events are estimated for luminosity $\mathcal{L}=3$ ab$^{-1}$. \label{tab:cut-flow}}
{\renewcommand\baselinestretch{1.25}\selectfont
\begin{tabular*}{\textwidth}{l @{\extracolsep{\fill}} %r
rrrrrr}
\hline
 %&  
 & \multicolumn{6}{c}{Selection Criteria}\\\cline{2-7} %\cline{2-7}
                                 %& \tcr{Gen. Cut} 
                                 & $\mathfrak{C}_1$ & $\mathfrak{C}_2$ & $\mathfrak{C}_3$ & $\mathfrak{C}_4$ & $\mathfrak{C}_5$ & $\mathfrak{C}_6$ \tabularnewline
\hline\hline
\multicolumn{7}{c}{Signal benchmarks} \\
\hline
$M_{B}=1200$ GeV, $M_{\Phi}=400$ GeV       %& $3499$     
& $2619$           & $1681$           & $1677$           & $1628$           & $1176$           & $1029$           \tabularnewline
$M_{B}=1500$ GeV, $M_{\Phi}=400$ GeV      %& $651$      
& $478$            & $333$            & $332$            & $321$            & $258$            & $223$            \tabularnewline
$M_{B}=1800$ GeV, $M_{\Phi}=700$ GeV      %& $141$      
& $102$            & $75$             & $75$             & $72$             & $60$             & $53$             \tabularnewline
\hline
\multicolumn{7}{c}{Background processes} \\
\hline
$t_{\ell}t_{h}~(+ 2j)$             %& $2800633$  
& $\num{2.12e6}$        & $\num{1.86e6}$        & $\num{1.74e6}$        & $\num{1.40e6}$        & $\num{4.84e5}$         & $\num{2.78e5}$         \tabularnewline
$W_{\ell}~(+ 2j)$            %& $1841915$  
& $\num{1.45e6}$        & $\num{1.35e6}$        & $\num{7.82e5}$         & $\num{1.57e5}$         & $75886$          & $40337$          \tabularnewline
$t_{\ell}t_{\ell}~(+ 2j)$          %& $84856$    
& $31460$          & $28244$          & $23377$          & $18490$          & $8833$           & $5519$           \tabularnewline
$t_{h/\ell}W_{\ell/h} ~(+ 1j)$                  %& $86397$    
& $63778$          & $53199$          & $43691$          & $36235$          & $9693$           & $4708$           \tabularnewline
$Z_{\ell} ~(+ 2j)$        
& $33771$          & $31786$          & $22847$          & $5324$           & $3247$           & $1962$           \tabularnewline         
$W_{\ell} Z_h ~(+ 1j)$                  %& $21530$    
& $17206$          & $16209$          & $10070$          & $7350$           & $3290$           & $1670$           \tabularnewline
$t_{\ell} t_{h}Z_h ~(+ 1j)$            %& $8957$     
& $6830$           & $6084$           & $5871$           & $4908$           & $2017$           & $1348$           \tabularnewline
$ W_{\ell}W_{h}~(+ 1j)$ %& $58786$    
& $44568$          & $38566$          & $25330$          & $8709$           & $2427$           & $1325$           \tabularnewline
$t_{\ell} t_{h}H ~(+ 1j)$            %& $4411$     
& $3360$           & $2976$           & $2907$           & $2740$           & $1219$           & $853$           \tabularnewline
$t_{\ell} + b/j$                            %& $12898$    
& $10830$          & $9324$           & $4357$           & $3311$           & $962$            & $608$            \tabularnewline
$t_{h}t_{\ell/h}W_{h/\ell}~(+ 1j)$      %& $667$      
& $556$            & $509$            & $501$            & $445$            & $113$            & $79$             \tabularnewline
\hline 
\multicolumn{2}{c}{ } & \multicolumn{4}{r}{Total number of background events:} & $\num{3.36e5}$\\
\hline\hline
\end{tabular*}}
\end{centering}
\end{table*}
%%%%%%%%%%%%%%%%%%%%%%%%%%%%%%%%%%%%%%%%%%

%%%%%%%%%%%%%%%%%%%%%%%%%%%%%%%%%%%%%%%%%%
\subsection{Kinematic features of the signal}
\noindent 
Since the selection cuts in $\mathfrak{C}_1$--$\mathfrak{C}_6$ are pretty strong, the surviving signal and background events look topologically very similar. Hence, we use a DNN  on the events passing $\mathfrak{C}_1$--$\mathfrak{C}_6$ to isolate the signal.  The network is trained on different kinematic distributions of the signal and background events. Each selected event has some well-defined objects---the high-$p_T$ lepton, the three AK-4 jets, the $b$-tagged jet, the fatjet, and missing $E^{\text{miss}}_T$  (since the lepton in the signal comes from the decay of a $W$-boson). We feed the network the following kinematic properties of these objects:
%%%%%%%%%%%%%%%%%%%%%%%%%%%%%%%%%%%%%%%%%%
\begin{figure*}
\centering
\captionsetup[subfigure]{labelformat=empty}
\subfloat[\quad\quad(a)]{\includegraphics[width=0.29\textwidth]{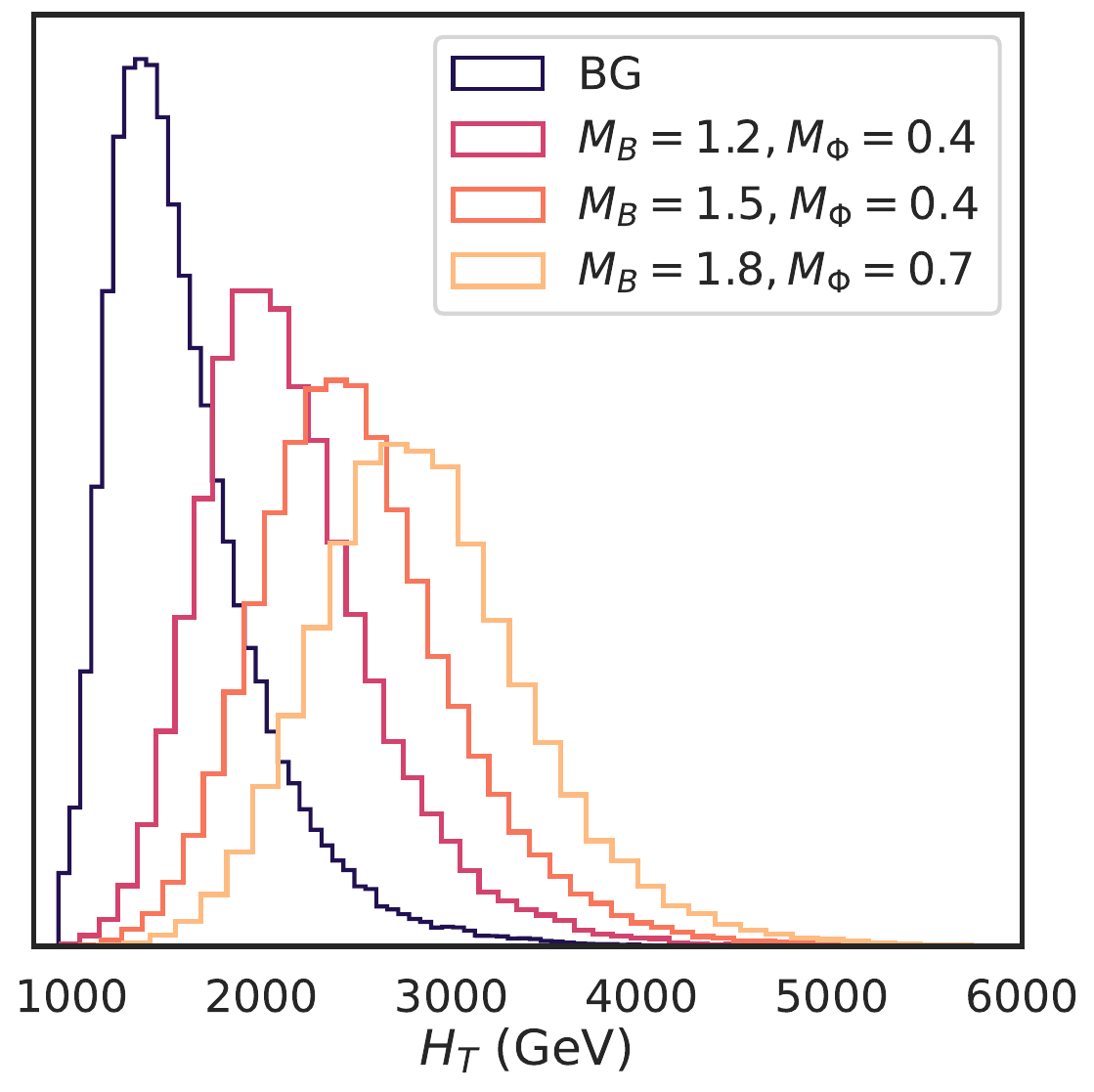}\label{fig:featDistA}}\quad\quad
\subfloat[\quad\quad(b)]{\includegraphics[width=0.29\textwidth]{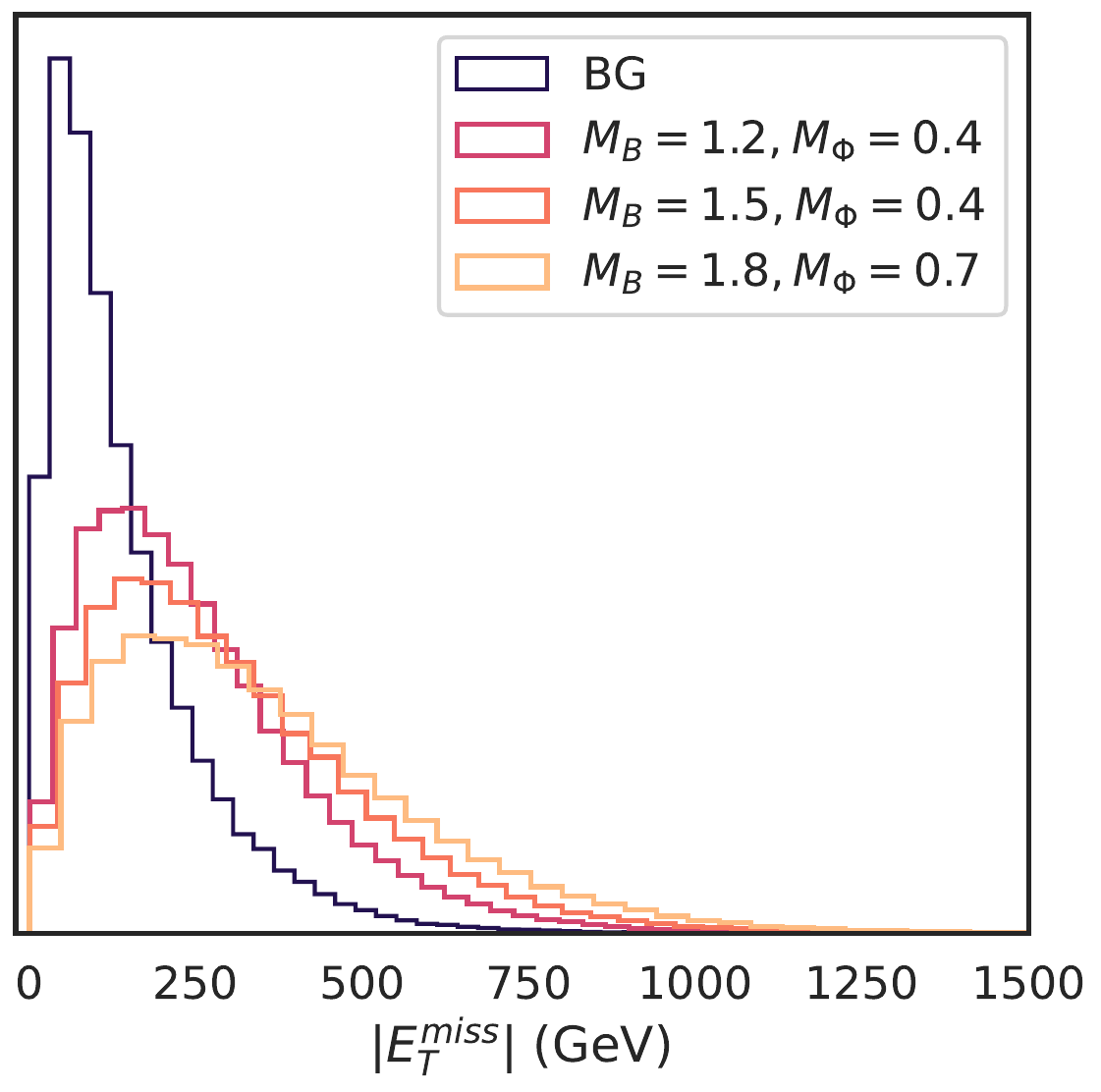}\label{fig:featDistB}}\quad\quad
\subfloat[\quad\quad(c)]{\includegraphics[width=0.29\textwidth]{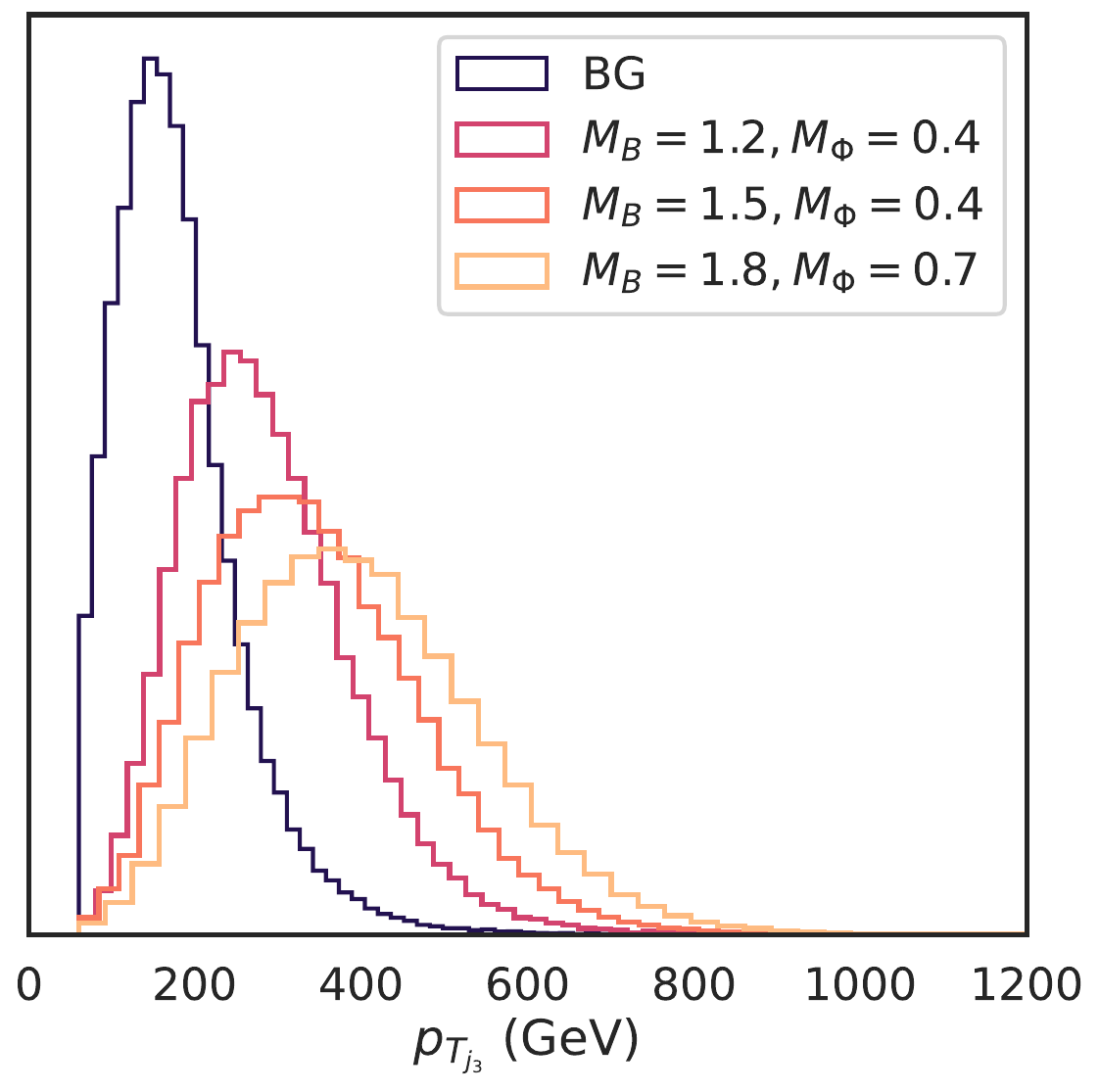}\label{fig:featDistC}}\\
\subfloat[\quad\quad(d)]{\includegraphics[width=0.29\textwidth]{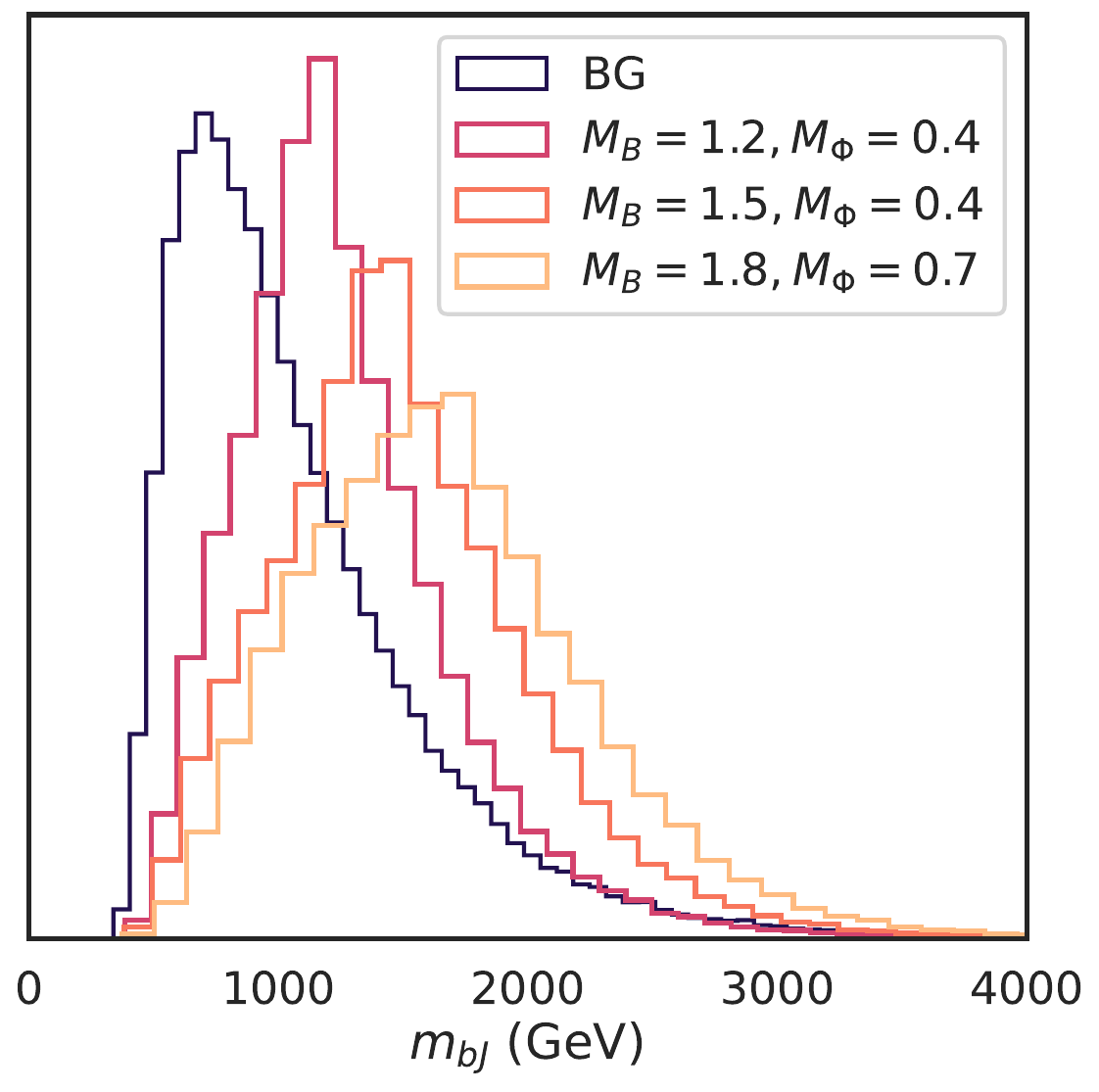}\label{fig:featDistD}}\quad\quad
\subfloat[\quad\quad(e)]{\includegraphics[width=0.29\textwidth]{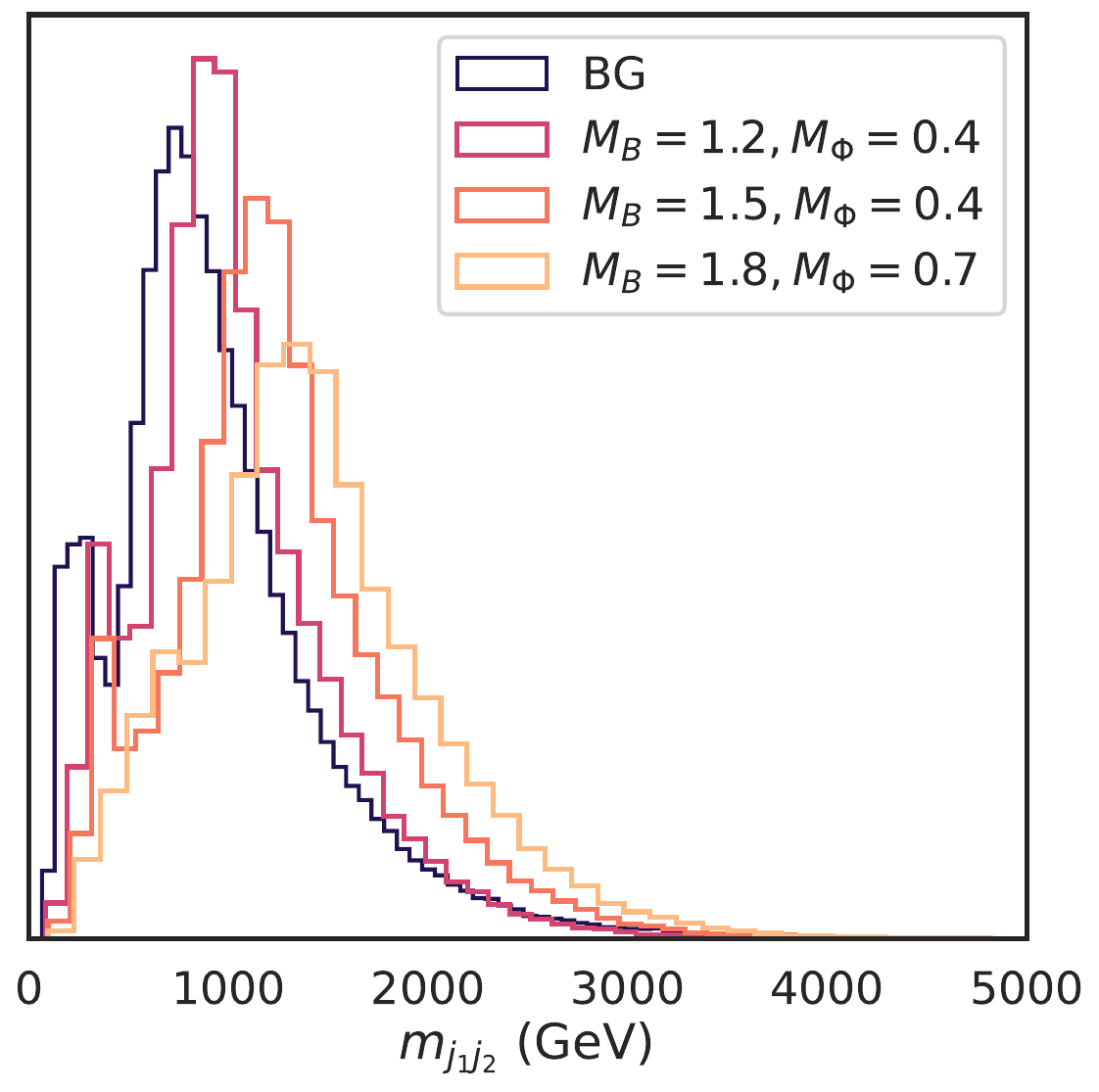}\label{fig:featDistE}}\quad\quad
\subfloat[\quad\quad(f)]{\includegraphics[width=0.29\textwidth]{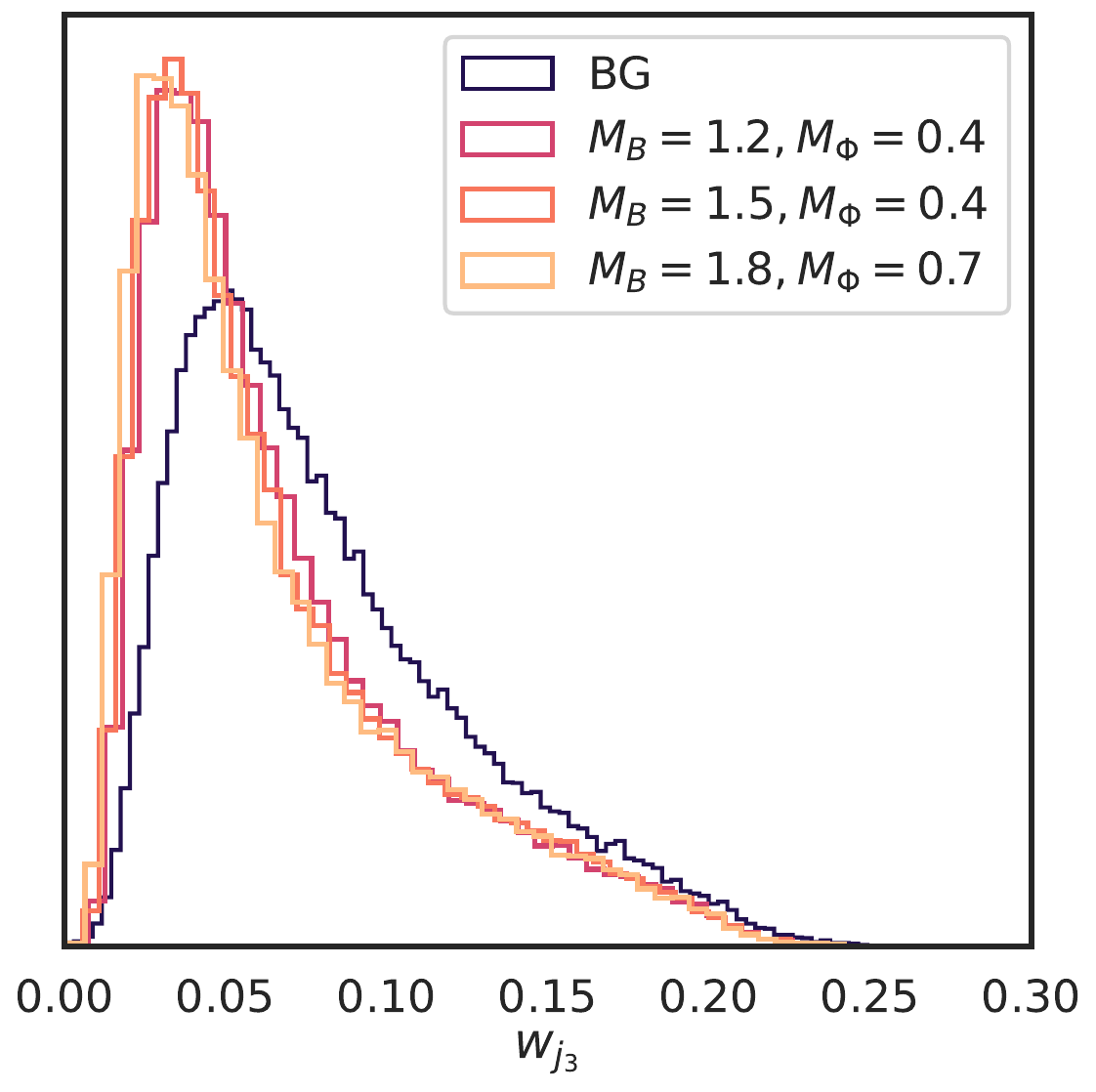}\label{fig:featDistF}}\\
\subfloat[\quad\quad(g)]{\includegraphics[width=0.29\textwidth]{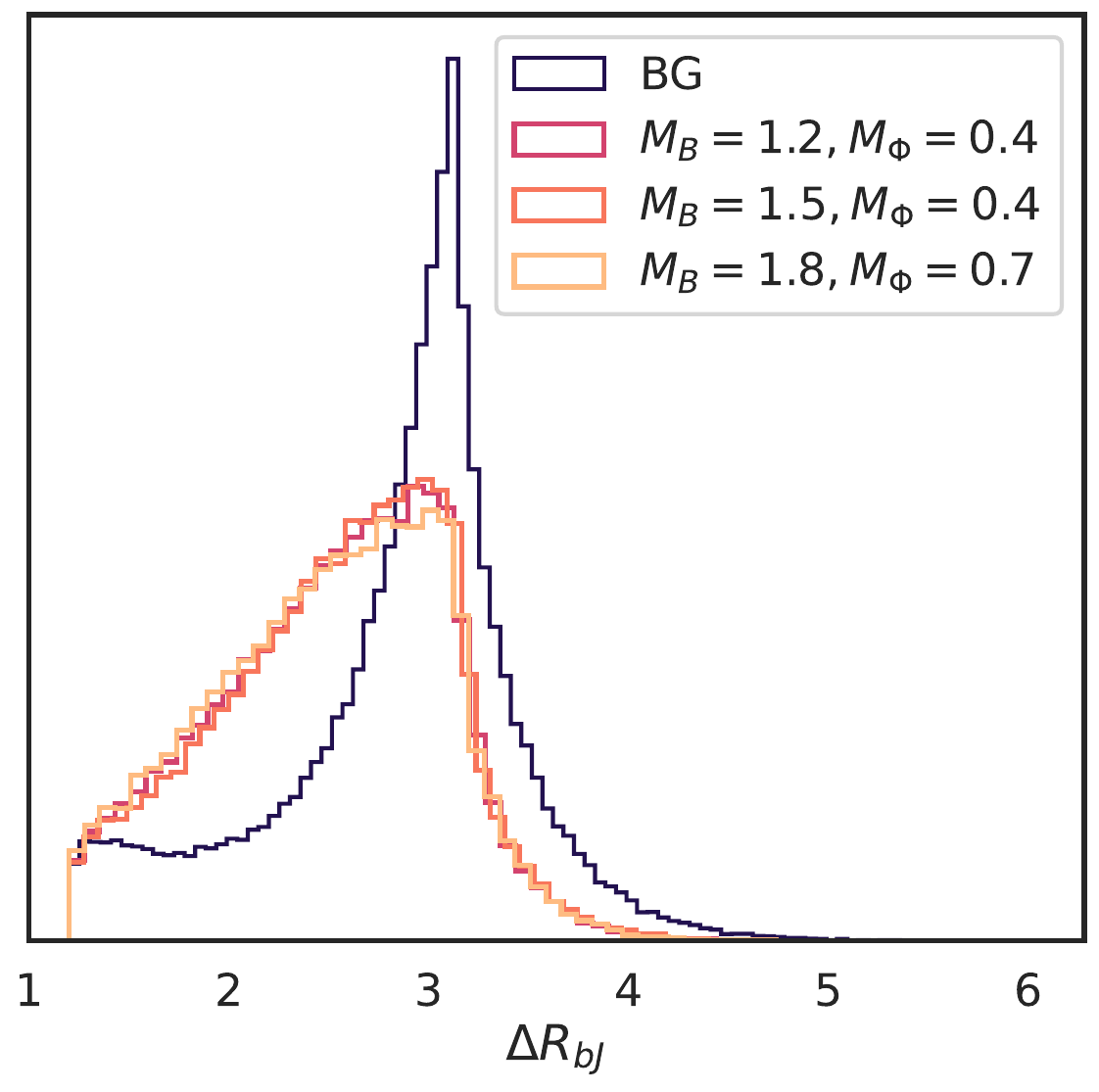}\label{fig:featDistG}}\quad\quad
\subfloat[\quad\quad(h)]{\includegraphics[width=0.29\textwidth]{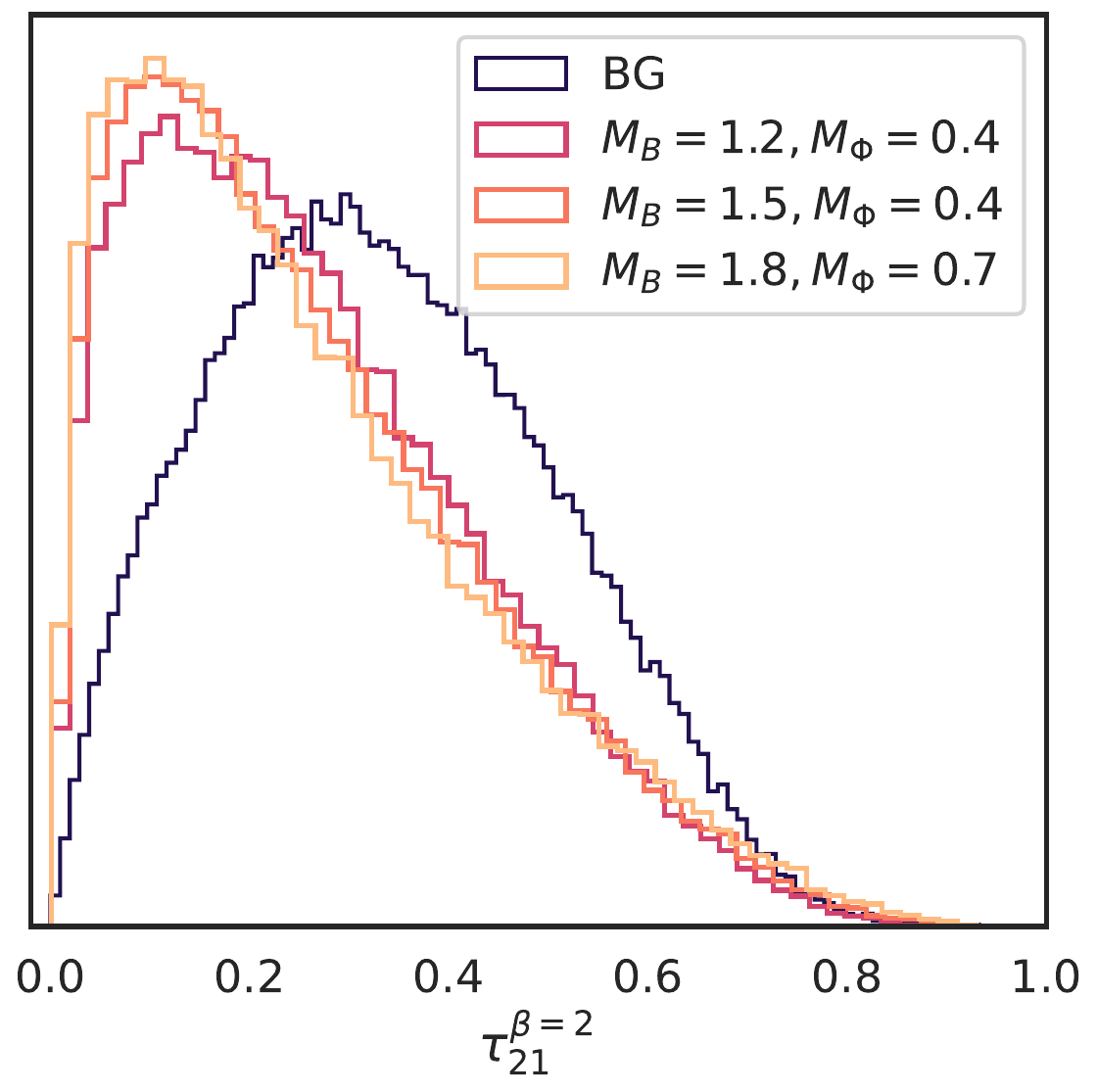}\label{fig:featDistH}}\quad\quad
\subfloat[\quad\quad(i)]{\includegraphics[width=0.29\textwidth]{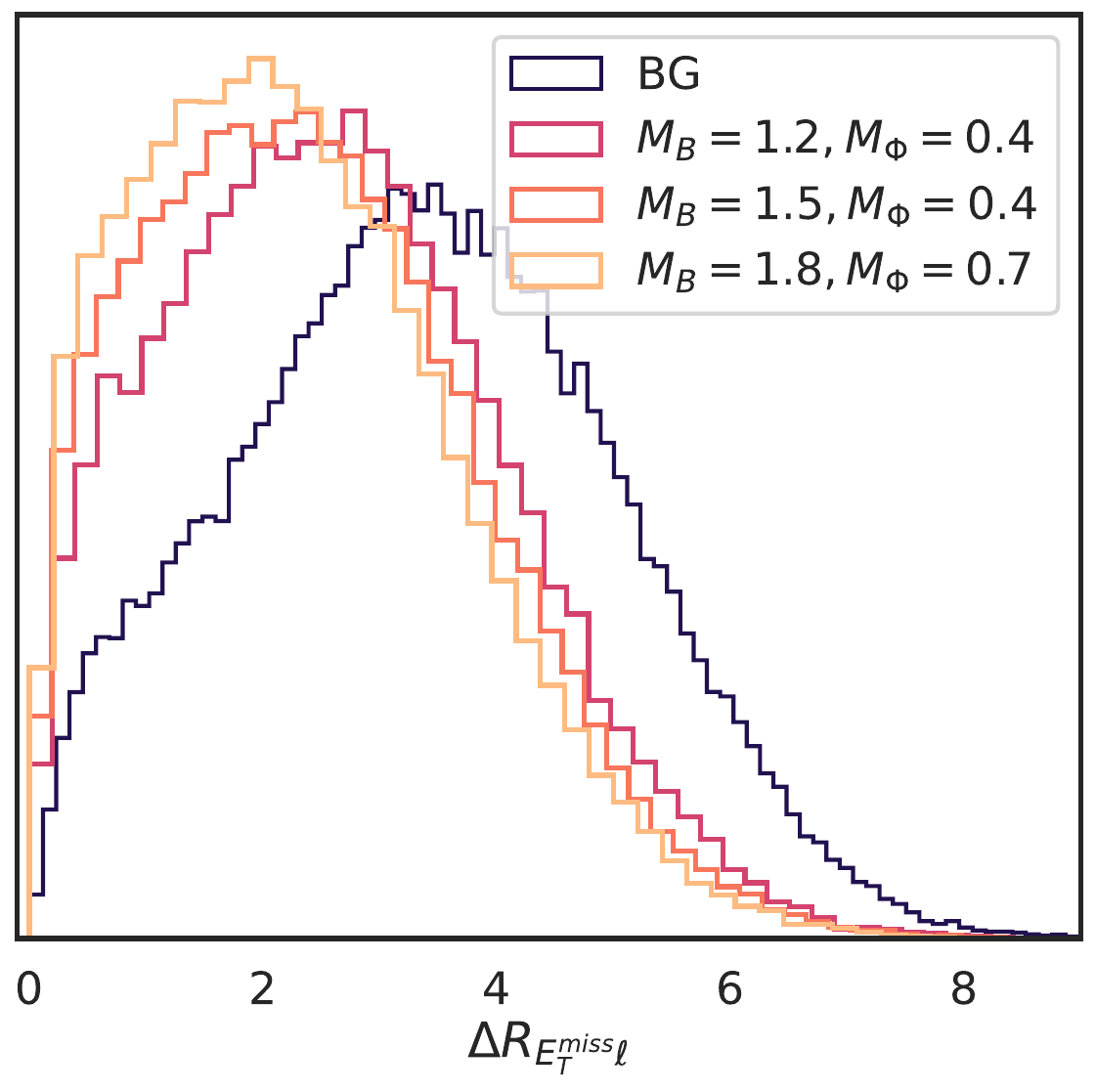}\label{fig:featDistI}}\\
\caption{Normalised density plots of select features for signal and background at various signal benchmarks  ($M_{\Phi}$ and $M_{B}$ are in TeV).}\label{fig:featDist}
\end{figure*}
%%%%%%%%%%%%%%%%%%%%%%%%%%%%%%%%%%%%%%%%%%
\begin{enumerate}
    \item \emph{Basic variables:} For each identified object, we consider the transverse momentum $(p_T)$. The scalar $H_T$ of the event and missing energy is also considered. The set of kinematic variables chosen is $\left\{H_T, |E^{\text{miss}}_T|, p_{T_\ell}, p_{T_{j1}}, p_{T_{j2}}, p_{T_{j3}}, p_{T_b}, p_{T_J}\right\}$.
    
    \item \emph{Jet-substructure variables:} For the fatjet, we calculate the $n$-subjettiness ratios $(n=1,2,3)$ for multiple $\beta$ values $(\beta = 1.0, 2.0)$ to take the prongness of $J$ into account. The set used is $\left\{ \tau_{21}^{\beta=1},\tau_{21}^{\beta=2}, \tau_{32}^{\beta=1}, \tau_{32}^{\beta=2}\right\}$.
    
    \item \emph{Distance in the $\eta-\phi$ plane:} We calculate the separation between two objects as $\Delta R_{ij} = \sqrt{\Delta \Phi_{ij}^2 + \Delta \eta_{ij}^2}$. We choose all possible pairs from the reconstructed objects and calculate the distance between them.
    
    \item \emph{Invariant masses of objects and their combinations:} We consider the masses of hadronic objects and the invariant masses of combinations of $2$ or $3$ objects. The set of (invariant) mass variables is  $\left\{m_{i'}, m_{ij}, m_{ijk} \right\}$, where $i'$ denotes an reconstructed hadronic object and each of $i$,$j$, and $k$ represent any reconstructed object.
    
    \item \emph{Girth/width of hadronic objects:} The girth (width) of a hadronic object is the $p_T$-weighted average distance of the constituents of the jet from the jet axis~\cite{Behr:2015oqq}. We also consider the higher-order central moments---variance, skewness and kurtosis of the distribution. The set of girth (width) related variables considered is $\left\{ g_{i'}, \text{Skew}[i'], \text{Kurt}[i'], \text{Var}[i']\right\}$, where $i'$ denotes a reconstructed hadronic object and Skew, Kurt, Var stand for the skewness, kurtosis, and the variance of the $p_T$-weighted distribution of the constituents of the hadronic object.
\end{enumerate}
We note that some of these variables can be correlated; for example, from the selection criteria, we expect that one of the leading three AK4 jets will be identical to the $b$ jet in some events since the $b$ jet is also an AK4 jet. In Fig.~\ref{fig:featDist}, we show the distributions for a subset of variables where the separation between the signal and the background distributions is clear. The distributions are shown for three benchmark parameter choices: (a) $M_{B} = 1.4$ TeV, $M_{\Phi} = 0.2$ TeV,  (b) $M_{B} = 1.5$ TeV, $M_{\Phi} = 0.5$ TeV, and (c) $M_{B} = 1.8$ TeV, $M_{\Phi} = 0.9$ TeV. We choose the benchmarks across the signal mass range to understand the key trends. The total background distribution is generated by combining events from different background processes in proportion to their cross sections.

The boosts of the identified objects in the signal are significantly higher than those in the background. This trend is evident from the distribution of transverse momentum of the sub-subleading jet, $j_3$ [Fig.~\ref{fig:featDistC}] where the signal peaks appear at higher values than the SM processes. Scalar $H_T$ [Fig.~\ref{fig:featDistA}] and missing transverse energy [Fig.~\ref{fig:featDistB}] distributions also indicate that the signal is well separated in the variables defined on the transverse plane. Fig.~\ref{fig:featDistD} and~\ref{fig:featDistE} show two invariant masses constructed from the identified objects---the invariant mass of the $b$ jet-fatjet pair $(m_{bJ})$ and the invariant mass of the leading two AK-4 jets $(m_{j_1 j_2})$, respectively. We expect the $m_{bJ}$ to reconstruct the $B$ mass most of the time; we see that peaks of the signal distributions appear close to the benchmark $B$ masses. Fig.~\ref{fig:featDistF} shows the girth of the sub-subleading jet.  The distributions for the signal benchmarks are almost identical and separated from the background distribution. As the boost of a jet increases, it becomes more collimated. Since the jets from the signal process are significantly more boosted than those from the background, the girth of the background peaks at a higher value than the signal points. This trend is broadly true for the girth distributions of all hadronic objects. The separations between the fatjet and the $b$ jet, shown in Fig.~\ref{fig:featDistG}, indicate that the fatjet and the selected $b$ jet are mostly back to back. In the signal, the $b$ jet from the $B\to b\Phi$ decay has higher $p_T$ than the one coming from the $B\to tW$ branch, but the $b$ tagging efficiency falls considerably as the $p_T$ jet increases. Hence, the $b$ jet closest to the fatjet is less likely to be $b$ tagged. In the case of the background, the fatjet is more often the hadronic top quark from the semileptonic $t\bar t$ process, where the $b$ jet comes from the leptonic top. The $n$-subjettiness ratio $(\tau_{21}^{\beta=2})$ in Fig.~\ref{fig:featDistH} shows that the signal distributions mostly have a two-pronged fatjet whereas the background fatjet is more likely to be a top quark. The separation between the missing energy and the lepton $(\Delta R_{E_{T}^{\text{miss}}\ell})$ shown in Fig.~\ref{fig:featDistI} indicate how boosted the leptonic $W$ is---the signal distributions peak at a lower value compared to the background distribution since, generally, the $W$'s in the signal have more boost. As the mass of the $B$ quark increases, the peak shifts to slightly lower values indicating a more collimated $W$.

%%%%%%%%%%%%%%%%%%%%%%%%%%%%%%%%%%%%%%%%%%
\subsection{Inclusive Search for (Pair-produced) $B$ in the $B \to tW$ Channel}\label{sec:IncMode}
\noindent 
We also estimate the reach of the monoleptonic channel in an inclusive scenario with the same DNN model. We demand that a pair-produced $B$ event should have at least one $W$ boson and at least one top quark, either decaying leptonically,
\begin{equation}\label{eq:InclSignal}
    pp \to B{B} \to \left\{ \begin{array}{l} \left(t W\right)\,\left({b}\Phi\right) \\ \left(t W\right)\,\left({t}W\right) \\ \left(t W\right)\,\left({b}H\right) \\ \left(t W\right)\,\left({b}Z\right) \end{array}\right.. 
\end{equation}
We set the selection criteria for the inclusive signal to be the same as in Sec.~\ref{set:CutsExcl} except for the fatjet-specific ones---here, we look for boosted $W$ bosons and tag them as $2$-pronged fatjets. We modify the cuts $\mathfrak{C}_5, \mathfrak{C}_6$ in Sec.~\ref{set:CutsExcl} as
\begin{itemize}
    \item[$\mathfrak{C}_5$:] \emph{Atleast 1 fatjet (J) with $R=0.6$ and $p_T > 300$ GeV, with $M_{J} \in \left[40,120\right]$}.
    \item[$\mathfrak{C}_6$:] \emph{$\Delta R_{b J} > 0.6$}
\end{itemize}
As mentioned earlier, the fatjets are clustered using the anti-$k_T$ algorithm; we have tuned the fatjet parameters to efficiently identify boosted $W$ jets.

%%%%%%%%%%%%%%%%%%%%%%%%%%%%%%%%%%%%%%%%%%
\section{Deep Neural Network Model}\label{sec:dnn}
\noindent
NNs consist of a series of perceptron blocks with a non-linear activation function. A perceptron block is simply a linear transformation of the input vector. NNs are motivated by the human brain, where a biological neuron is modelled by a perceptron block and the neuron interactions are modelled with non-linear activation functions of the form: 
\begin{align}
    f(\mathbf{x}) = \sigma(\mathbf{W} \mathbf{x} + \mathbf{b}),
\end{align}
where $\sigma$ is an activation function (like sigmoid, ReLU, etc.). Deep NNs can effectively approximate any real continuous function~\cite{HornikEtAl89} provided the dimensions of the hidden layers are sufficiently large.

We use a standard DNN architecture to boost the significance of the $B$ signal at the HL-LHC. The selected network is composed of two linear layers (of $128$ dimension) with Mish Activation~\cite{2019arXiv190808681M} and Batch Norm~\cite{2015arXiv150203167I}. Additionally, Dropout~\cite{JMLR:v15:srivastava14a} with a dropout probability of $0.2$ and L2 weight decay of $10^{-4}$ are used to regularise the training. We obtain the NN architecture by a grid search over the hyperparameters and train the network with the \texttt{AdamW} optimizer~\cite{2017arXiv171105101L}. 

Classification tasks generally minimise a cross-entropy loss between classes to get the best performance. However, a simple cross-entropy loss does not account for the differences in cross sections of the various processes because each event is weighed equally for training the network. Since we are considering multiple processes with different cross sections, the higher the cross section of a process, the higher should be the penalty for misclassifying events from that process. In other words, the weight of each event should scale positively with the cross section of the generating process.\footnote{Weighing samples according to their true distribution is known to yield better-performing classifiers in other domains~\cite{10.1117/12.2624729}.} Similarly, if we feed the network a large number of events from a single process, the weight of an individual event should decrease with that number---there are more examples for the classifier to learn from. Therefore, we use a weighted form of the cross-entropy loss to train the network where the weight $\omega$ of an event of a particular process $p_i$ is given as,
\begin{equation}
    \omega_{p_i} = \sqrt{\frac{\sigma_{p_i}\ {L}}{\mathcal{N}_{p_i}}}.
\end{equation}
Here, $\sigma_{p_i}$ denotes the cross section of the process $p_i$, $\mc N_{p_i}$ is for the number of events fed to the network, and $L=3$ ab$^{-1}$ is the experimental luminosity. The function form of the weight (i.e., the square root) is determined empirically, for the weights perform better with the square root than without. Disregarding the effect of discretisation when calculating the signal significance, we find a good correlation between loss and significance. The models are implemented using \textsc{Pytorch} ~\cite{NEURIPS2019_bdbca288} and trained on a \texttt{Nvidia GTX 1080Ti GPU}, even though it trains fairly quickly on a CPU too. 
%%%%%%%%%%%%%%%%%%%%%%%%%%%%%%%%%%%%%%%%%%
\begin{figure}[t!]
    \centering
    \includegraphics[scale=0.7]{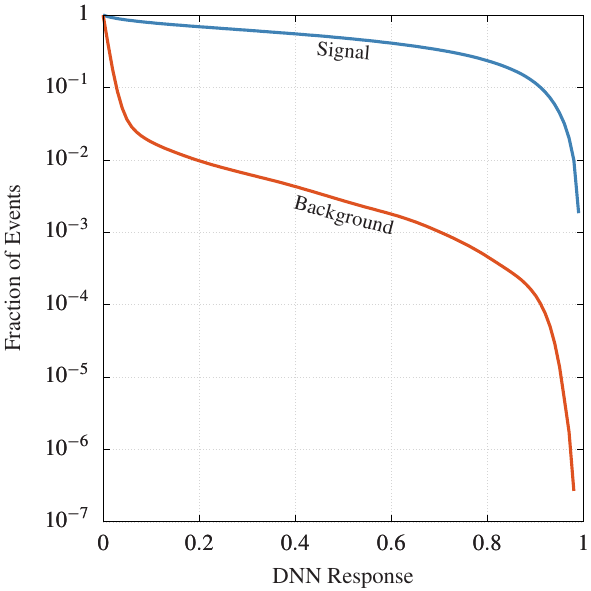}
    \caption{The fractions of signal and background events surviving the network-response thresholds for $M_B = 1.5$ TeV and $M_\Phi = 0.4$ TeV.}
    \label{fig:DNN_NsNb}
\end{figure}
%%%%%%%%%%%%%%%%%%%%%%%%%%%%%%%%%%%%%%%%%%
%%%%%%%%%%%%%%%%%%%%%%%%%%%%%%%%%%%%%%%%%%
\begin{figure*}
\captionsetup[subfigure]{labelformat=empty}
    \centering
    \subfloat[\quad\quad(a)]{{\includegraphics[height=5.5cm]{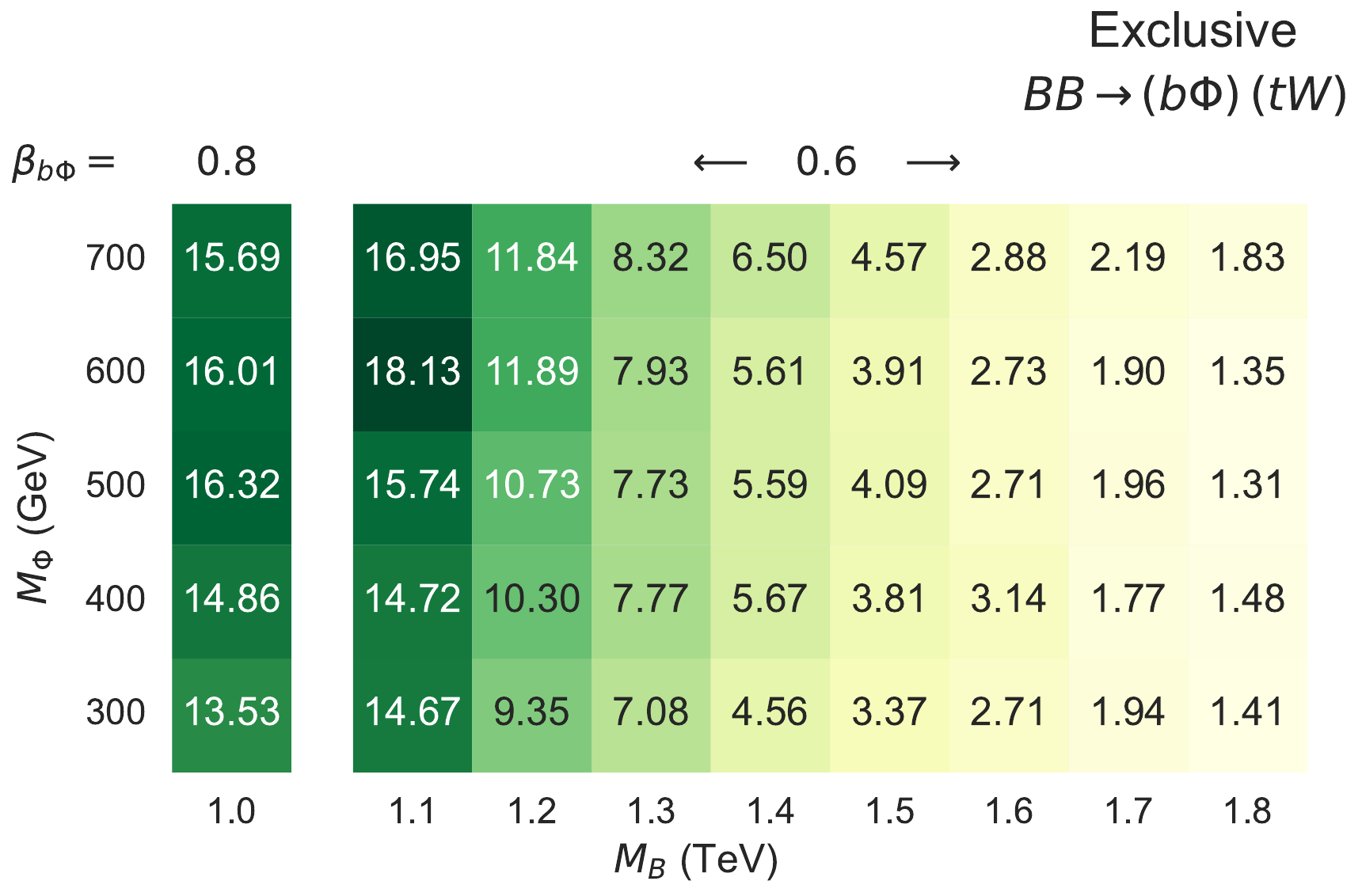}}\label{fig:sigGridA}}\hfill
    \subfloat[\quad\quad(b)]{{\includegraphics[height=5.5cm]{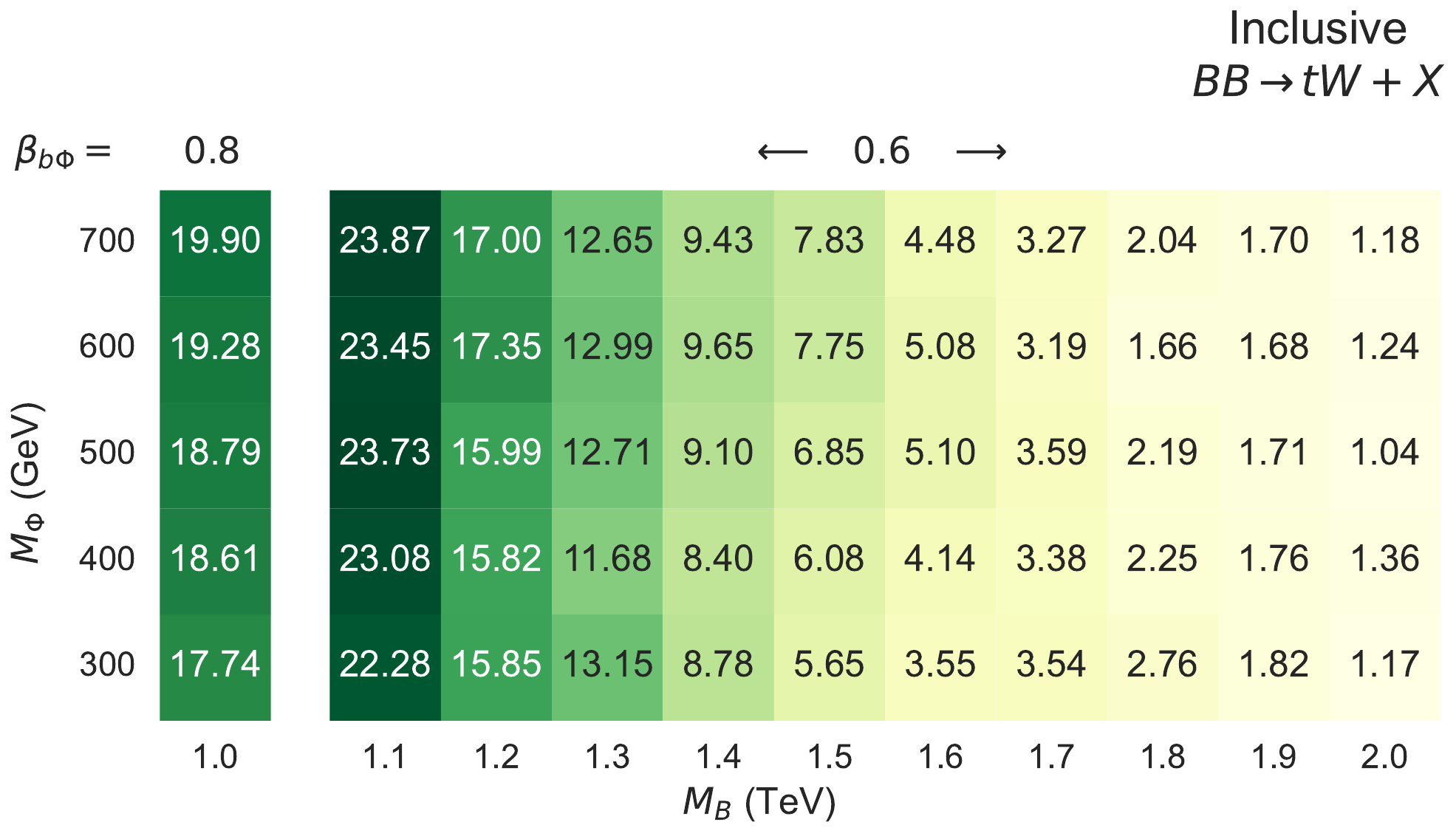}}\label{fig:sigGridB}}
    \caption{Signal significances predicted by the DNN model for various combinations of $M_{B}$ and $M_{\Phi}$ for the (a) exclusive signal and (b) inclusive signal at the HL-LHC ($L=3$ ab$^{-1}$). BR$(B\to b\Phi)$ for each parameter point is shown on the top. Significance estimates for $M_{B}=1$ TeV are presented separately since the rescaled LHC limits (Fig.~\ref{fig:b2_LHClimits}) rule out a TeV $B$ for BR$(B\to b\Phi)\lesssim 0.7$. We assume BR$(\Phi \to jj) = 1.0$.}
    \label{fig:sigGrid}
%\end{figure*}
%%%%%%%%%%%%%%%%%%%%%%%%%%%%%%%%%%%%%%%%%%%
%%%%%%%%%%%%%%%%%%%%%%%%%%%%%%%%%%%%%%%%%%%
%\begin{figure*}
    \centering
\captionsetup[subfigure]{labelformat=empty}
\subfloat[\quad\quad(a)]{{\includegraphics[width=0.375\textwidth]{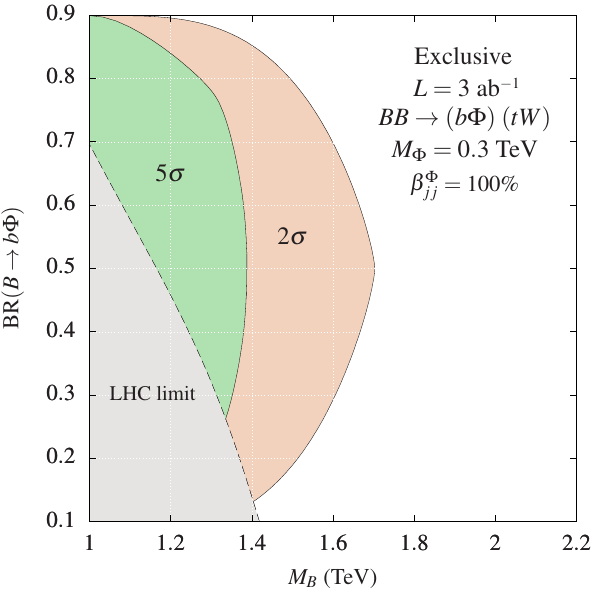}}\label{fig:brContoursA}}\hspace{1cm}
\subfloat[\quad\quad(b)]{{\includegraphics[width=0.375\textwidth]{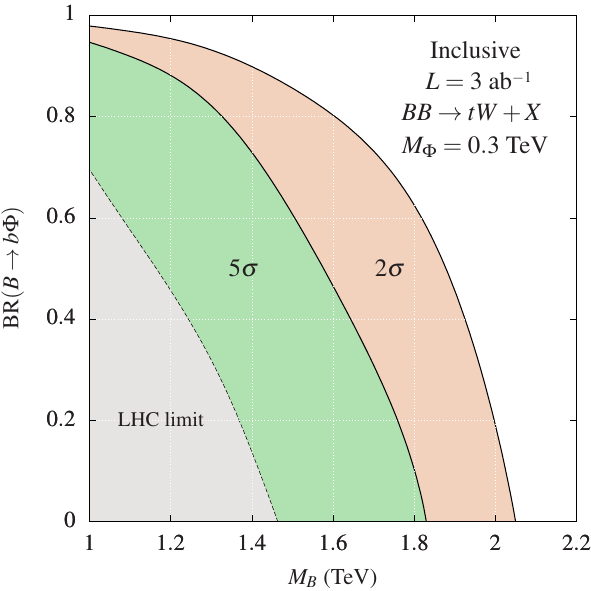}}\label{fig:brContoursC}}\\
\subfloat[\quad\quad(c)]{{\includegraphics[width=0.375\textwidth]{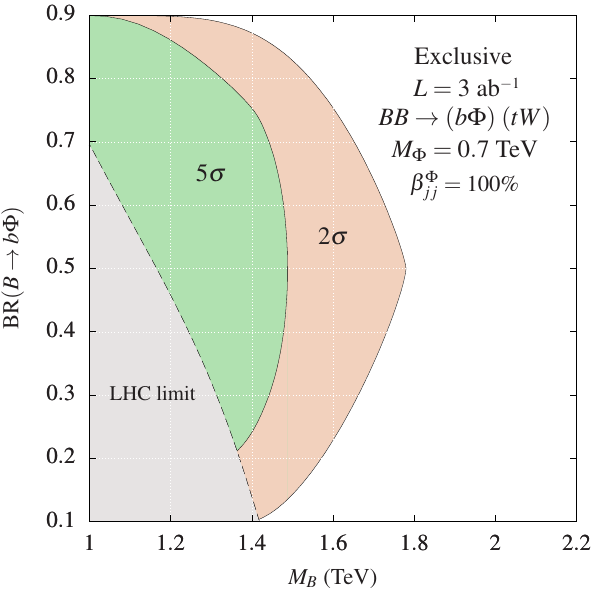}}\label{fig:brContoursB}}\hspace{1cm}
\subfloat[\quad\quad(d)]{{\includegraphics[width=0.375\textwidth]{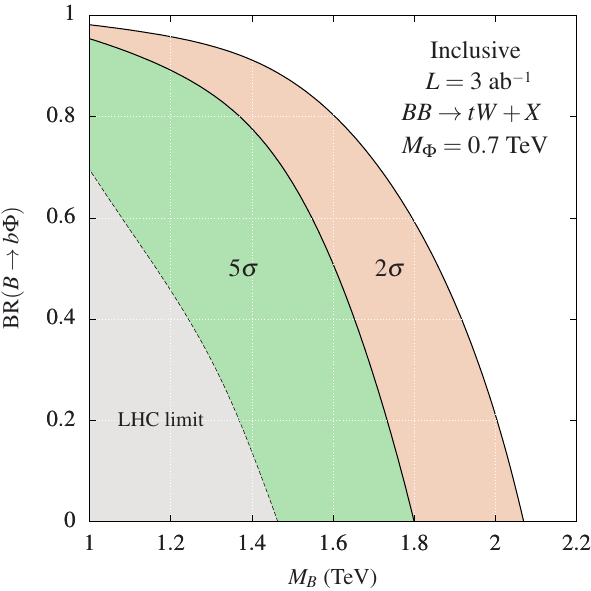}}\label{fig:brContoursD}}\\
\caption{HL-LHC reach plots for the exclusive [$BB \to b\Phi\,tW$; on the left panel, (a) and (c)] and inclusive  [$BB \to tW+X$; on the right panel, (b) and (d)] modes. The grey areas are ruled out by LHC limits shown in Fig.~\ref{fig:b2_LHClimits}.}
\label{fig:brContours}
\end{figure*}
%%%%%%%%%%%%%%%%%%%%%%%%%%%%%%%%%%%%%%%%%%%

While estimating the signal significance, we perform a scan across the network response to achieve the best performance. To tune the network, we pick the signal benchmark point $(M_{B},M_{\Phi}) = (1.5, 0.4)$ TeV which is near the centre of the parameter space considered. We use \textsc{Weights and Biases}~\cite{wandb} which provides tools and neat visualisations for metrics required to identify the best-performing  model. We use its grid-search feature to scan over a multi-dimensional grid of hyperparameters. We search over the following set of hyperparameters: (a) the number of hidden layers of the network, (b) the number of nodes in each hidden layer, (c) the learning rate, (d) L2 regularisation co-efficient, and (e) the dropout rate. Regularisation and dropout are ways to prevent the network from overfitting to the training data. The L2 regularisation shrinks the weights of the layers and prevents the network from learning complex functions that are usually highly sensitive to noise and prone to overfitting, whereas dropout reduces the network's dependence on a particular set of variables by randomly `dropping' (making them unavailable) certain variables while training the network. 

We select the best-performing model with the least complexity, i.e., the least number of hidden layers and nodes in each layer as the final candidate for our analysis. The performance metric is the signal significance ($\mc Z$ score),
\begin{equation}
    \label{eq:Zscore}
    \mathcal{Z} = \sqrt{2\left(N_S + N_B\right)\ln\left(\frac{N_S+N_B}{N_B}\right) - 2N_S}
\end{equation}
where $N_S$ and $N_B$ are the numbers of signal and background events allowed by the network at $3$ ab$^{-1}$.
We scan over the network response---a value between $0$ (ideal background) and $1$ (ideal signal)---on the validation dataset against the predicted significance and pick the network response with a high $\mc Z$ score as the threshold for classifying a signal event. Fig.~\ref{fig:DNN_NsNb} shows that as we increase the network response, i.e., demand a more stringent classification, $N_S$ decreases smoothly but $N_B$ falls drastically.

We also identify and discuss the input features deemed important by the neural network using a popular NN interpretability technique, Integrated Gradients~\cite{integrated_gradients}, in Appendix~\ref{app:ig-interpret}.

%%%%%%%%%%%%%%%%%%%%%%%%%%%%%%%%%%%%%%%%%%
\section{Results}\label{sec:results}
\noindent We present the projected signal significance at the HL-LHC predicted by the DNN model in the $M_{B}-M_{\Phi}$ plane and also as a function of the BR in the new decay mode in Figs.~\ref{fig:sigGrid}--\ref{fig:brContours}. The results for the exclusive $\left(BB\to tW\,b\Phi\right)$ and inclusive channel $\left(BB\to tW+X\right)$ are presented separately on the left and right panels, respectively. 

%%%%%%%%%%%%%%%%%%%%%%%%%%%%%%%%%%%%%%%%%%
\subsection{Exclusive mode}
\noindent Fig.~\ref{fig:sigGridA} shows the significance predictions from the DNN model at various mass points for $\bt_{b\Phi}=0.6$ (except for the $M_B=1$ TeV case, where the LHC data exclude $\bt_{b\Phi}\lesssim0.7$). In Figs.~\ref{fig:brContoursA} and ~\ref{fig:brContoursB}, we show the  $5\sigma$ (discovery) and  $2\sigma$ ($\sim$ exclusion) contours on the $\beta_{b\Phi}$--$M_{B}$ plane for two benchmark masses of $\Phi$, $M_{\Phi}=300$, $700$ GeV. The strong demands on the mass of the fatjet and its $p_T$ ($\mathfrak C_5$) prefer a boosted heavy $M_\Phi$ in the signal. We roughly see this behaviour in Fig.~\ref{fig:sigGridA}. The shaded regions in Figs.~\ref{fig:brContoursA} and~\ref{fig:brContoursB} are symmetric about $\bt_{b\Phi}=0.5$ since the signal yield scales as $\beta_{b\Phi} \left(1 - \beta_{b\Phi}\right)$ [Eq.~\eqref{eq:brEq}]. The grey regions are excluded by the rescaled LHC limits (shown in Fig.~\ref{fig:b2_LHClimits}) for the singlet $B$ model. 

%%%%%%%%%%%%%%%%%%%%%%%%%%%%%%%%%%%%%%%%%%
\subsection{Inclusive mode}
 \noindent We see from Fig.~\ref{fig:sigGridB} that the inclusive mode has better significance scores than the exclusive mode for the same parameter choice. The improvement is also visible in the $5\sigma$ and $2\sigma$  contours shown in Figs.~\ref{fig:brContoursC} and \ref{fig:brContoursD}. The branching factor determining the signal yield in the $tW$-inclusive mode is given by [see Eq~(40) of Ref.~\cite{Bhardwaj:2022nko}]
\begin{align}
\mc B_{tW}^{incl} = \beta_{tW}^2  + 2\sum_{X \neq tW}\beta_{tW}\beta_{X}= \beta_{tW}\left(2 - \beta_{tW}\right).\label{eq:inclufactor}
\end{align}
In our case, we can express the branching factor in terms of $\bt_{b\Phi}$ in the singlet $B$ model by combining the above expression with Eq.~\eqref{eq:brEq} as
\begin{align}
        \mc B_{tW}^{incl} &= \beta_{tW}\left(2 - \beta_{tW}\right) \nonumber \\
                          &\approx \left(\frac{1-\beta_{b\Phi}}{2}\right) \left(2 - \frac{1-\beta_{b\Phi}}{2}\right) \nonumber \\
                          &= \frac{1}{4}\left(1-\beta_{b\Phi}\right) \left( 3+\beta_{b\Phi}\right). \label{eq:BRtWbPhi}
\end{align}
Hence, the branching factor should be maximum at $\beta_{b\Phi}=0$ and zero at $\beta_{b\Phi}= 1$. These trends are seen in Figs.~\ref{fig:brContoursC} and~\ref{fig:brContoursD}. 

There are mainly two reasons for the better significance scores in the inclusive mode than in the exclusive mode. First, since $\mc B_{tW}^{incl} > \bt_{b\Phi}\left(1-\beta_{b\Phi}\right)$ for  $\beta_{b\Phi}<1$, the signal is larger in the inclusive mode.  Second, we find that the network is slightly better at improving signal significance in the inclusive mode. However, one should remember the difference in their interpretations when comparing the two results. Theoretically, the inclusive mode is insensitive to the nature of $\Phi$ (i.e., this particular decay of $B$), unlike the exclusive mode (where the $\Phi$ jet-motivated selection criteria create additional restrictions). The results should not change if the $B$ decayed not to $\Phi$ but through a different exotic decay mode (with different kinematics or topology) or multiple new decay modes. However, even though we make our initial selection criteria insensitive to $\Phi$, the final network may still learn of its features through the pair-produced $B$ events (where one $B$ quark decays through the $\Phi$ mode in a fraction of events) we use for training. Hence, our results for the inclusive mode should be taken as indicative.

%%%%%%%%%%%%%%%%%%%%%%%%%%%%%%%%%%%%%%%%%%
\begin{figure*} %[htp]
    \centering
    \includegraphics[width=\textwidth]{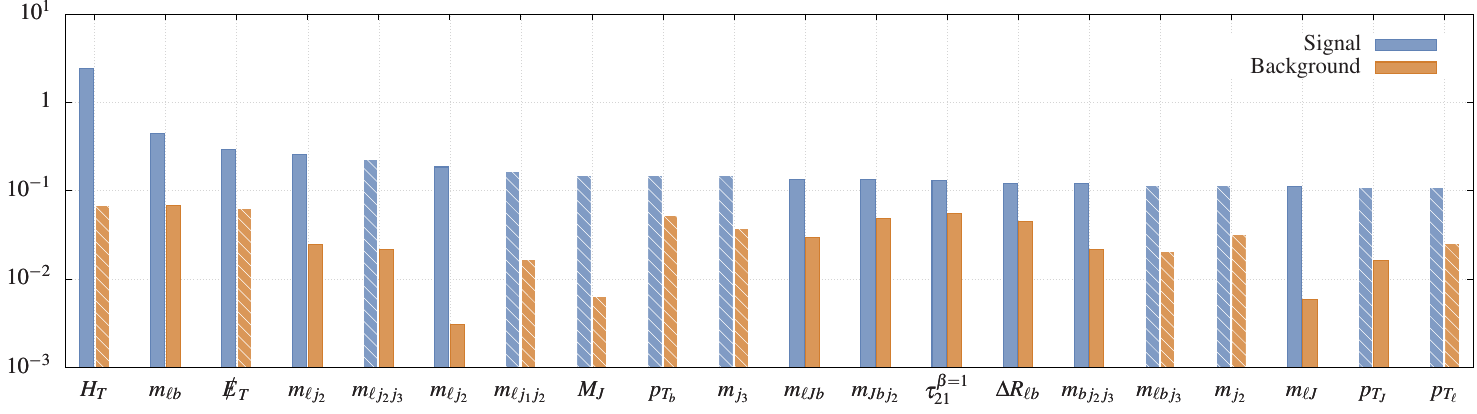}
    \caption{Top twenty features sorted by their importance for signal-class prediction from the DNN model. We define the baseline by averaging over the background events [Eq.~\eqref{eq:ExpectedGradients}]. Blue bars denote the importance of a feature to predict the signal class and orange bars denote the importance to predict the background class. The negative importance values are shown with patterned bars.}
    \label{fig:DNNvarImp}
\end{figure*}
%%%%%%%%%%%%%%%%%%%%%%%%%%%%%%%%%%%%%%%%%%
%%%%%%%%%%%%%%%%%%%%%%%%%%%%%%%%%%%%%%%%%%
\begin{figure*}
\captionsetup[subfigure]{labelformat=empty}
\centering
\subfloat[\quad\quad\quad(a)]{\includegraphics[width=0.375\textwidth]{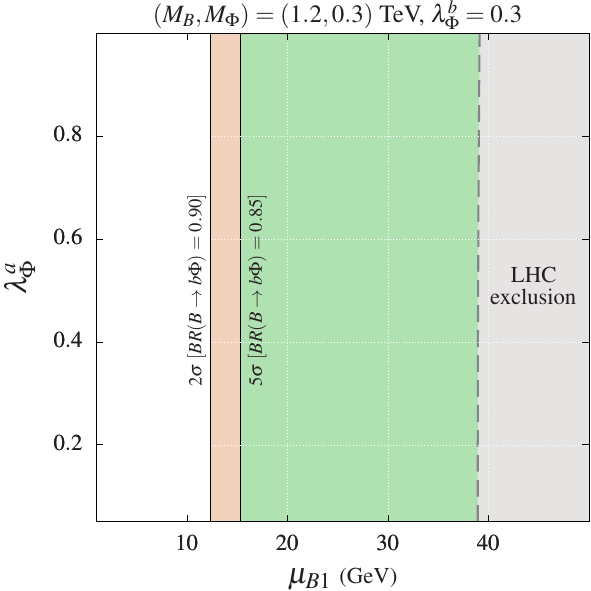}\label{fig:ModelContours_lambdaB}}\hspace{1cm}
\subfloat[\quad\quad\quad(b)]{\includegraphics[width=0.375\textwidth]{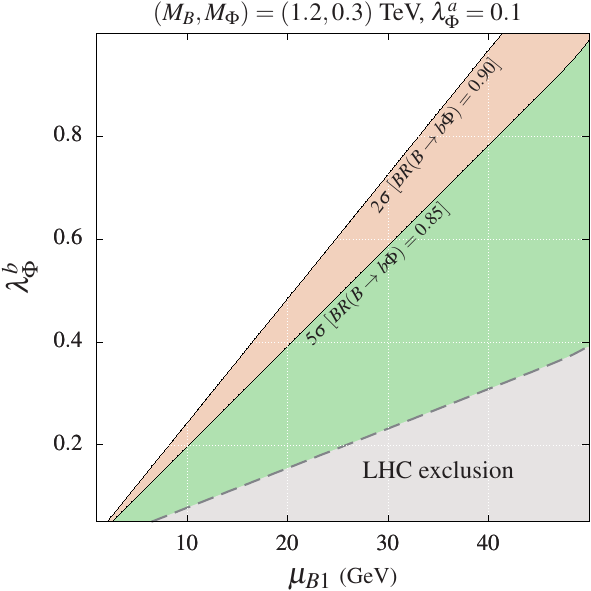}\label{fig:ModelContours_lambA}}
\caption{The $5\sigma$ and $2\sigma$ significance contours from Fig.~\ref{fig:brContoursA} in terms of the model parameters: $\m_{B1} = \om_B v/\sqrt 2$, the off-diagonal term in the mass matrix [defined in Eq.~\eqref{eq:massmat}] and the couplings, $\lambda_a$ and $\lambda_b$ [defined in Eq.~\eqref{eq:lambdas}]. }
\label{fig:ModelContours}
\end{figure*}
%%%%%%%%%%%%%%%%%%%%%%%%%%%%%%%%%%%%%%%%%%

%%%%%%%%%%%%%%%%%%%%%%%%%%%%%%%%%%%%%%%%%%
\section{Conclusions}\label{sec:conclu}
\noindent
As a followup to Refs~\cite{Bhardwaj:2022nko, Bhardwaj:2022wfz}, in this paper, we studied the discovery/exclusion prospects of the vectorlike weak-singlet $B$ quark in the presence of a lighter weak-singlet spinless state $\Phi$ at the HL-LHC. The singlet $\Phi$, which otherwise shares no direct couplings with the SM quarks, couples with the third-generation quarks when they mix with VLQs (which directly couple with $\Phi$) after EWSB. Hence, after EWSB, $\Phi$ can have tree-level decays to third-generation quark pairs. It can also decay into a pair of gluons (or, in fewer cases, into pairs of other SM bosons) through quark loops. In Ref.~\cite{Bhardwaj:2022nko}, we mapped the possibilities to explore such setups at the LHC. We showed that for a VLQ to decay into $\Phi$ and the $\Phi$ to decay to a pair of gluons or third-generation quark(s), no fine-tuning of the parameters is needed.

For this study, we focused on the $B$ quark pair production channel in particular where one $B$ quark decays to a $b$ quark and a $\Phi$, and the other decays to a top quark and a $W$ boson. This is an exclusive signature of the singlet $B$ model~\cite{Bhardwaj:2022nko}, which, when the $W$ decays leptonically, offers a leptonic handle at the LHC. Otherwise, the fully hadronic final states from the $B$ quark pair production are considerably more challenging to probe compared to the semileptonic case or the case of the $T$ quark in the presence of $\Phi$. We postpone the analysis of the fully hadronic channel to a future publication.

Since the $\Phi$ dominantly decays to a pair of $b$ quarks or gluons in the singlet $B$ model~\cite{Bhardwaj:2022nko}, we considered a set of signal selection criteria that is agnostic of $\Phi$ decay modes, which helped us to explore a large part of the parameter space of the singlet $B$ model. We further used a DNN model with a weighted cross-entropy loss trained on a large set of kinematic variables to isolate the signal for various benchmark masses of $B$ and $\Phi$. We presented the statistical significances obtained by the DNN model along with the discovery and exclusion regions as functions of BR in the new mode. We used the same DNN model to study the prospects of the monoleptonic channel in an inclusive scenario also. In this case, one of the pair-produced $B$ quarks produces a $W$ boson that decays leptonically.  

Our estimations showed the HL-LHC could probe a large region of the parameter space with the monolepton signals. Specifically, with $3$ ab$^{-1}$ of integrated luminosity, the discovery reach for a $B$ quark could go up to $\sim1.5$ TeV and $\sim1.8$ TeV in the exclusive and inclusive monolepton channels, respectively; whereas the $2\sigma$ limits go up to $\sim 1.8$ and $\sim 2.1$ TeV, respectively. 

%%%%%%%%%%%%%%%%%%%%%%%%%%%%%%%%%%%%%%%%%%
\section*{Model Files}
\noindent 
The {\sc Universal FeynRules Output}~\cite{Degrande:2011ua} files used in our analysis are available at \url{https://github.com/rsrchtsm/vectorlikequarks/} under the name {\tt SingBplusPhi}. 

%%%%%%%%%%%%%%%%%%%%%%%%%%%%%%%%%%%%%%%%%%
\acknowledgements
\noindent
C. N. acknowledges the Department of Science and Technology (DST)-Inspire for his fellowship.

\appendix
%%%%%%%%%%%%%%%%%%%%%%%%%%%%%%%%%%%%%%%%%%
\begin{figure}
    \centering
    \includegraphics[width=0.375\textwidth]{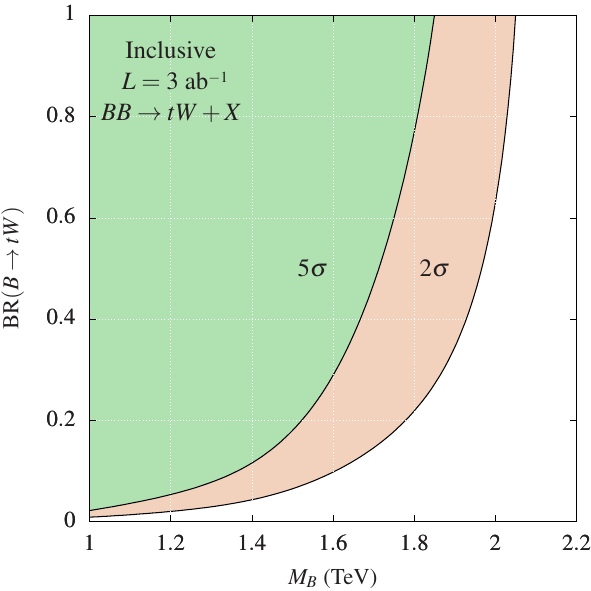}
    \caption{The inclusive $5\sigma$ and $2\sigma$ significance contours in terms of BR$(B \to tW)$.\label{fig:Inc_tW_res}}
\end{figure}
%%%%%%%%%%%%%%%%%%%%%%%%%%%%%%%%%%%%%%%%%% 
%%%%%%%%%%%%%%%%%%%%%%%%%%%%%%%%%%%%%%%%%%
\section{Interpretation with Integrated Gradients}\label{app:ig-interpret}
\noindent
Integrated Gradients~\cite{integrated_gradients} is a method to interpret the relative importance of the input features in the prediction of a machine-learning (ML) model. It is based on the idea that the contribution of every input feature to the model's prediction can be approximated by integrating the gradient of the model's output with respect to that input feature along the straight-line path between a baseline input and the input. %(the null vector, $\Vec{0}$). 
Mathematically, if we have a ML model with input $\mathbf x$ and output $\mathbf y$, the contribution of the $i^\text{th}$ input feature to the model's prediction can be approximately expressed as
\begin{equation}
    \mathfrak G_i^{IG}(f, \mathbf x, \mathbf x') = (x_i - x'_i)  \int_{\alpha = 0}^1 \frac{\delta f\left[\mathbf x' + \alpha (\mathbf x - \mathbf x')\right]}{\delta x_i} d \alpha
\end{equation}
where $\mathbf x$ and $\mathbf x'$ represent the features of the input and the baseline input, respectively. The integral is taken from $\alpha = 0$ to $\alpha = 1$ and $(\mathbf x - \mathbf x^\prime)$ represents the vector from the baseline to the input,  $\delta f\left[\mathbf x' + \alpha (\mathbf x - \mathbf x')\right]/\delta x_i$ is the gradient of the model's output with respect to the $i^\text{th}$ input feature.
By calculating the integrated gradients and averaging over the events, we can estimate the relative importance of each input feature. The sign of the integrated gradient for an input feature indicates whether the feature had a positive or negative impact on the model's prediction. The magnitude indicates the relative scale of impact the feature had on the prediction compared to other features.

In our context, we want to study how signal events differ from background events. Hence, we could take a background event as the baseline. However, there is no obvious single baseline but the fact that $\phi_{IG}$ should be insensitive to the properties of the baseline (i.e., the baseline should be informationless). Here, we average over multiple baselines~\cite{sturmfels2020visualizing,smoothgrad}. This leads to a modified form:
\begin{equation}
    \mathfrak G_i(f,\mathbf x) = \int_{\mathbf x'} \mathfrak G_i^{IG}(f, \mathbf x, \mathbf x') \times p_D(\mathbf x')\,d\mathbf x' \label{eq:ExpectedGradients}
\end{equation}
where the baseline $x'$ is integrated over a probability distribution on baselines $D$. We take the distribution of the background events  (in proportion to the cross sections of the originating processes) as the distribution of the baselines. Eq.~\eqref{eq:ExpectedGradients} can be seen as computing the expectation over the set of baselines, which can be approximated from a few samples using the Monte Carlo estimation technique.

Fig.~\ref{fig:DNNvarImp} shows that $H_T$ has the highest importance in pushing the model to predict the signal class. This is expected since there is a considerable separation in $H_T$ [Fig.~\ref{fig:featDist}]. Furthermore, $\slashed{E}_T$ and $m_{\ell b}$ also seem to be essential to the network for predicting the signal class. We also see that from the left towards right, the importance falls very quickly implying that the first few variables are enough to provide good separation. While there are some negative importance scores, they are close to $0$; we can think of them as tiny correction terms to the final classifier output rather than essential contributors. The feature attribution falls only slightly for the ones following the top $10$ due to the weight-decay regularisation used while training the network. For example, the $40^{\text{th}}$ most-important variable, $m_{\slashed{E}_T j_3}$, has a feature attribution of $\approx 0.056$.\medskip

\section{Interpreting $B\to b\Phi$ prospects}

\noindent
In Figs.~\ref{fig:sigGrid} and~\ref{fig:brContours}, we present the significance contours in terms of $M_B$, $M_\Phi$ and $\bt_{b\Phi}$ for better interpretability in a wide class of models.  Apart from the masses of $B$ and $\Phi$, our model is characterised by three parameters: the $b$--$B$ mixing parameter $\m_{B1}=\omega_B v/\sqrt2$ in the mass matrix $\mc M$ [Eq.~\eqref{eq:massmat}] and the couplings, $\lambda_a$ and $\lambda_b$ [Eq.~\eqref{eq:lambdas}]. It is straightforward to translate our results in terms of the model parameters with the decay width expressions in Ref.~\cite{Bhardwaj:2022nko}. For example, we can reinterpret Fig.~\ref{fig:brContoursA} in terms of the model parameters  for $M_B=1.2$ TeV as shown in Fig.~\ref{fig:ModelContours}.

In our model, the exclusive signal scales as $2\,\bt_{b\Phi}\,\lt(1-\bt_{b\Phi}\rt)/2$. Hence, in a generic model where $\bt_{tW}$ is independent of $\bt_{b\Phi}$ and $\bt^\Phi_{jj}< 1$, the signal strength will be scaled by $2\bt_{tW}\bt^\Phi_{jj}/\lt(1-\bt_{b\Phi}\rt)$. Since the $\mc Z$ score scales roughly linearly ($\approx N_S/\sqrt{N_B}$), the significance scores in Fig.~\ref{fig:sigGridA} will scale with the same factor. Reinterpreting the signal significance for the inclusive mode is even simpler since it essentially depends only on $\bt_{tW}$ [see Eq.~\eqref{eq:inclufactor}]. In this case, the signal strength and its significance scale by $4\beta_{tW}\left(2 - \beta_{tW}\right)/\left( 3-2\beta_{b\Phi} -\beta_{b\Phi}^2\right)$. We draw the inclusive $5\sg$ and $2\sg$ significance contours in terms of the $B\to t W$ branching ratio in Fig.~\ref{fig:Inc_tW_res}.

\def\bibfont{\small}
\bibliography{bPrime_MixedMode}{}
\bibliographystyle{JHEPcust}
\end{document}